\documentclass[10pt,conference]{IEEEtran}
\IEEEoverridecommandlockouts

\def\BibTeX{{\rm B\kern-.05em{\sc i\kern-.025em b}\kern-.08emT\kern-.1667em\lower.7ex\hbox{E}\kern-.125emX}}

\usepackage{makecell}
\usepackage{tabularx}
\usepackage{booktabs} 
\usepackage{array} 
\usepackage[most]{tcolorbox} 
\usepackage{balance}
\usepackage{multirow}
\usepackage{adjustbox}
\usepackage{listings}
\usepackage{xcolor}
\usepackage{graphicx}
\usepackage{setspace} 
\usepackage{comment}
\definecolor{deepgreen}{RGB}{30,80,30}

\newtcolorbox{conclusionbox}{
  colback=gray!10,
  colframe=gray!60,
  left=10pt,
  right=10pt,
  top=5pt,
  bottom=5pt,
  boxrule=1pt,
  arc=0pt,
  boxsep=5pt,
  breakable,
}

\newcommand{\find}[1]{
\begin{tcolorbox}[leftrule=1mm,toprule=0mm,bottomrule=0mm,left=1pt,right=2pt,top=2pt,bottom=2pt] 
#1
\end{tcolorbox}
}

\newcommand{\zp}[1]{\textcolor{black} {{#1}} }

\begin{document}
\title{Towards Better Answers: Automated Stack Overflow Post Updating}


\author{\IEEEauthorblockN{Yubo Mai}
\IEEEauthorblockA{
\textit{Zhejiang University}\\
Hangzhou, China \\
12021077@zju.edu.cn}
\and
\IEEEauthorblockN{Zhipeng Gao\textsuperscript{*}}
\IEEEauthorblockA{\textit{Shanghai Institute for Advanced Study of Zhejiang University}\\
Shanghai, China \\
zhipeng.gao@zju.edu.cn}
\and
\IEEEauthorblockN{Haoye Wang}
\IEEEauthorblockA{\textit{Hangzhou City University}\\
Hangzhou, China \\
wanghaoye@hzcu.edu.cn}
\and
\IEEEauthorblockN{Tingting Bi}
\IEEEauthorblockA{\textit{The University of Western Australia}\\
Perth, Australia \\
tingting.bi@uwa.edu.au}
\and
\IEEEauthorblockN{Xing Hu}
\IEEEauthorblockA{\textit{Zhejiang University}\\
Hangzhou, China \\
xinghu@zju.edu.cn}
\and
\IEEEauthorblockN{Xin Xia}
\IEEEauthorblockA{\textit{Zhejiang University}\\
Hangzhou, China \\
xin.xia@acm.org}
\and
\IEEEauthorblockN{Jianling Sun}
\IEEEauthorblockA{\textit{Zhejiang University}\\
Hangzhou, China \\
sunjl@zju.edu.cn}
\thanks{* This is the corresponding author}
}

\maketitle


\begin{abstract}
Utilizing code snippets on Stack Overflow (SO) is a common practice among developers for problem-solving. 
Although SO code snippets serve as valuable resources, it is important to acknowledge their imperfections, reusing problematic code snippets can lead to the introduction of suboptimal or buggy code into software projects. 
\textit{SO comments} often point out weaknesses of a post and provide valuable insights to improve the quality of answers, while SO comments are usually missed and/or ignored, leaving these problematic code snippets untouched. 
In this work, we first investigate the task of automatic SO posts updating based on their associated comments. 
We introduce a novel framework, named \textbf{\textsc{Soup}} (\textbf{\underline{S}}tack \textbf{\underline{O}}verflow \textbf{\underline{U}}pdator for \textbf{\underline{P}}ost) for this task. 
\textsc{Soup} addresses two key tasks: Valid Comment-Edit Prediction (VCP) and Automatic Post Updating (APU). 
We fine-tuned a large language model, CodeLlama, using low-rank adaptation techniques to complete the VCP task, and constructed a dataset containing 78k valid comment-edit pairs for the APU task. Subsequently, we tested the performance of multiple large language models on the APU task. Extensive experimental results show the promising performance of our model over a set of benchmarks. Moreover, we also perform an in-the-wild evaluation on Stack Overflow, we submitted 50 edits generated by our approach to Stack Overflow posts and 21 of them have been verified and accepted by SO maintainers, further proving the practical value of \textsc{Soup}.


\end{abstract}


\section{Introduction}
\label{sec:intro}

As one of the most popular programming Q\&A communities today, Stack Overflow (SO) offers a wealth of knowledge for solving various programming issues, which is scattered throughout its question and answer posts~\cite{gao2020generating, gao2023know}. 
However, not all SO posts are ``perfect'', many code snippets in the posts are suboptimal (e.g., needing code simplification or extension), problematic (e.g., using obsolete APIs or libraries), or even buggy (e.g., security flaws)~\cite{ragkhitwetsagul2019toxic}. 
Reusing these code snippets can mislead answer seekers and induce potential problems and/or bugs into their working code base, decreasing software quality and maintainability. 

\begin{figure}[h]
\centering
\includegraphics[width = 0.99\linewidth]{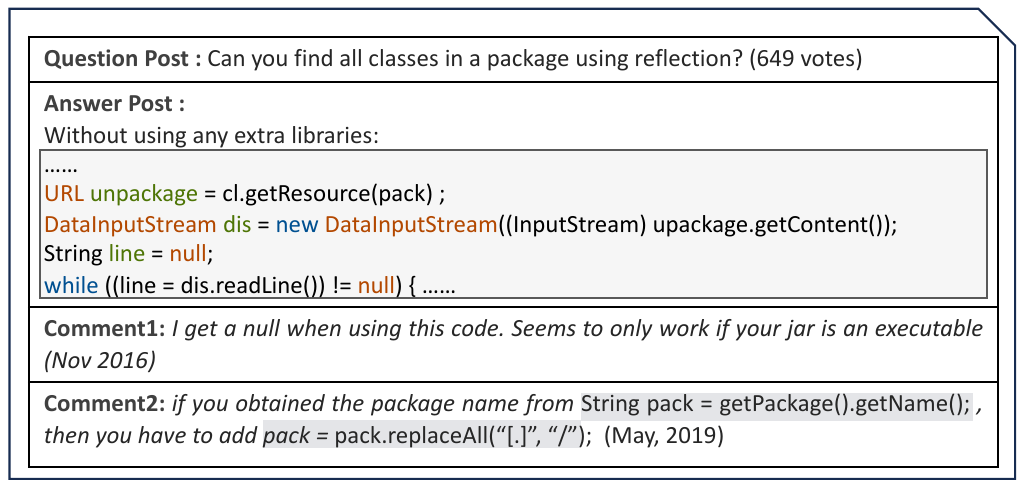}
\caption{Stack Overflow Post Example}
\vspace{-0.3cm}
\label{fig:intro}
\end{figure}

Fortunately, SO comments provide useful information to enhance the associated answers from a diverse range of perspectives, such as pointing out weaknesses and suggesting improvements. 
However, these SO comments can be easily overlooked or ignored (in contrast to answers), therefore, most comments are rarely integrated back into their associated answers~\cite{ragkhitwetsagul2019toxic, zhang2019empirical, nie2017data, gao2020technical}.  
Fig.~\ref{fig:intro} illustrates such an example, the most recent edits of this answer occurred in 2014, all subsequent comments (e.g., comment 1 \& 2) were overlooked and never integrated into answers. 
Moreover, since SO only displayed five comments to users, these newly posted comments can be hidden by SO and never be noticed.

Comments are the main channels for SO users to discuss and communicate potential problems in answer posts. 
Ideally, when a comment is posted to point out problem(s), it should trigger an update in corresponding answers (e.g., code snippets) and thus improve their quality. 
Nonetheless, because digesting comments and manually maintaining answers (after a long time being posted) are time-consuming and labor-intensive, users seldom update code snippets based on comments. 
As replied by an user in Stack Overflow~\cite{intro_so} when his/her obsolete answer was found, ``\textit{Feel free to update the answer yourself, if you like. I honestly would, but I don't have the time.}'' 
Therefore, it is highly desirable to have a tool that automatically updates SO code snippets according to the given comments; however, developing such a tool poses significant challenging, as outlined below: 
\begin{itemize}
    \item \textit{\textbf{High quality dataset}}. 
  Some previous works built datasets for SO post updates; for example, Tang et al.~\cite{tang2021using} first proposed a matching-based approach to map comments to their related edits, building a dataset containing 248k comment-edit pairs. 
    However, due to the limitations of their approach, the quality of their constructed dataset can not be guaranteed. 
    The recall of their method for identifying valid comment-edit pairs is only 12.6\%, while the precision is 57.1\%. 
    The relatively low-quality data greatly hinders the model's performance trained on it. To create such a dataset, it requires to extract valid comment-edit pairs from SO, i.e., a comment and its corresponding edit that addressed this comment. Currently, there is no high-quality dataset available for training models that automatically update posts.
    \item \textit{\textbf{Post updating }\textbf{models}.} Even Tang et al.~\cite{tang2021using} proposed the first comment-edit dataset, they ended up with creating the dataset without further utilizing it. 
    To the best of our knowledge, there has been no research delving into the automatic updating SO posts based on their comments. 
\end{itemize}

To address the aforementioned challenges, we define two key tasks as follows: (1) The \textbf{V}alid \textbf{C}omment-edit \textbf{P}rediction (VCP). 
For a given SO answer post, numerous comments can be added under the same post. However, not every comment leads to an edit.
In order to automatically construct a dataset that can be used to train models for automatic post updating, we defined this preliminary task. VCP is defined as predicting whether a comment-edit pair is valid, and a valid comment-edit pair is defined as: (i) the comment is relevant to the edit; (ii) the edit addresses/implements all the issues/suggestions raised in the comment; (iii) the edit does not make changes beyond what is mentioned in the comment. In this type of data, the updated code fully complies with the comments' suggestions, and therefore, can be used to train the APU model.
(2) The \textbf{A}utomatic \textbf{P}ost \textbf{U}pdating (APU). In Stack Overflow, an edit is associated with a pair of posts before and after modification. 
The APU task is defined as giving a post before modification and its associated comment, whether the post after modification can be correctly generated. 
Since developers are more interested in code snippets, this paper focuses only on the updating of code snippets in the posts.

Inspired by the great advancements and potential of Large Language Models (LLMs)~\cite{touvron2023llama, wang2024software, wang2022self, zhao2023survey, mai2024human, wang2024just, wang2024makes, gao2021automating, gao2024automating} in code generation~\cite{fried2022incoder, jiang2023selfevolve, lu2021codexglue, zheng2023codegeex, yan2023closer, dai2024mpcoder, xue2024selfpico}, in this work, we proposed a novel LLM-based framework, named \textbf{\textsc{Soup}} (\textbf{\underline{S}}tack \textbf{\underline{O}}verflow \textbf{\underline{U}}pdator for \textbf{\underline{P}}ost), to perform the VCP and APU tasks. 
For the VCP task, we first manually annotated 5K comment-edit pairs, we then fine-tuned a LLM for this task, and the trained model is denoted as \textsc{Soup}$_p$.  
Then the well-trained model \textsc{Soup}$_p$ is used to automatically create a high quality dataset consisting of 78K valid comment-edit pairs. 
This high-quality dataset is used for fine-tuning a LLM on the APU task, and the trained model is denoted as \textsc{Soup}$_u$. 
Our study aims to answer the following four research questions:

\begin{enumerate}
    \item \textbf{RQ1: How effective is our approach in predicting valid comment-edit pairs?} We compared our \textsc{Soup}$_p$ with the matching-based method proposed by Tang et al.~\cite{tang2021using}, the experimental findings demonstrate that our approach achieved an impressive 80.8\% precision and 74.0\% recall on the VCP task, marking a substantial improvement over Tang's method by 42\% and 487\% respectively. 
    \item \textbf{RQ2: How effective is our approach in automatically updating SO code snippets based on their comments?} 
    The experimental results show that our approach achieved a 25.6\% exact match rate, particularly excelling in solving improvement-type updates. 
    \item \textbf{RQ3: To what extent does the dataset influence model performance in the APU task?} We performed a cross-dataset evaluation between different models, the experimental results show that the model trained on our dataset significantly outperformed the same one trained on Tang's~\cite{tang2021using} proposed dataset. 
    \item \textbf{RQ4: How acceptable are our updated SO posts in real-world scenarios?} In this RQ, we conducted an in-the-wild evaluation to evaluate the effectiveness of \textsc{Soup} in practice. 
    We randomly sampled 50 unaddressed comments from SO, we then manually submitted \textsc{Soup} generated edits to these posts, 21 updates are accepted by SO maintainers. 
\end{enumerate}

Overall, our paper makes the following contributions:
\begin{itemize}
    \item We propose a novel \textbf{LLM-based framework}, {\sc Soup}, to accurately predict valid comment-edit pairs and autonomously update SO code snippets based on associated comments. \textsc{Soup} is expected to become a real-time updating tool for code forum websites, ultimately improving the code quality of such sites.
.
    \item We build a \textbf{high quality dataset} for SO post updating, which contains 78,317 valid comment-edit pairs for Java, the experimental results show that our constructed high quality dataset can significantly improve model's performance on the APU task. 
    \item We extensively evaluate our approach by submitting {\sc Soup} suggested edits to actual SO posts. 
    21 edits are accepted, verifying the practical value of our approach. 
    \item Our contribution also lies in the provision of both the source code and dataset~\cite{replicate} for \textsc{Soup}, enabling fellow researchers to replicate our findings and explore their own concepts with ease.
\end{itemize}



\section{Motivation}
\label{sec:def}

\begin{figure}[]
\centering
\includegraphics[width = 0.99\linewidth]{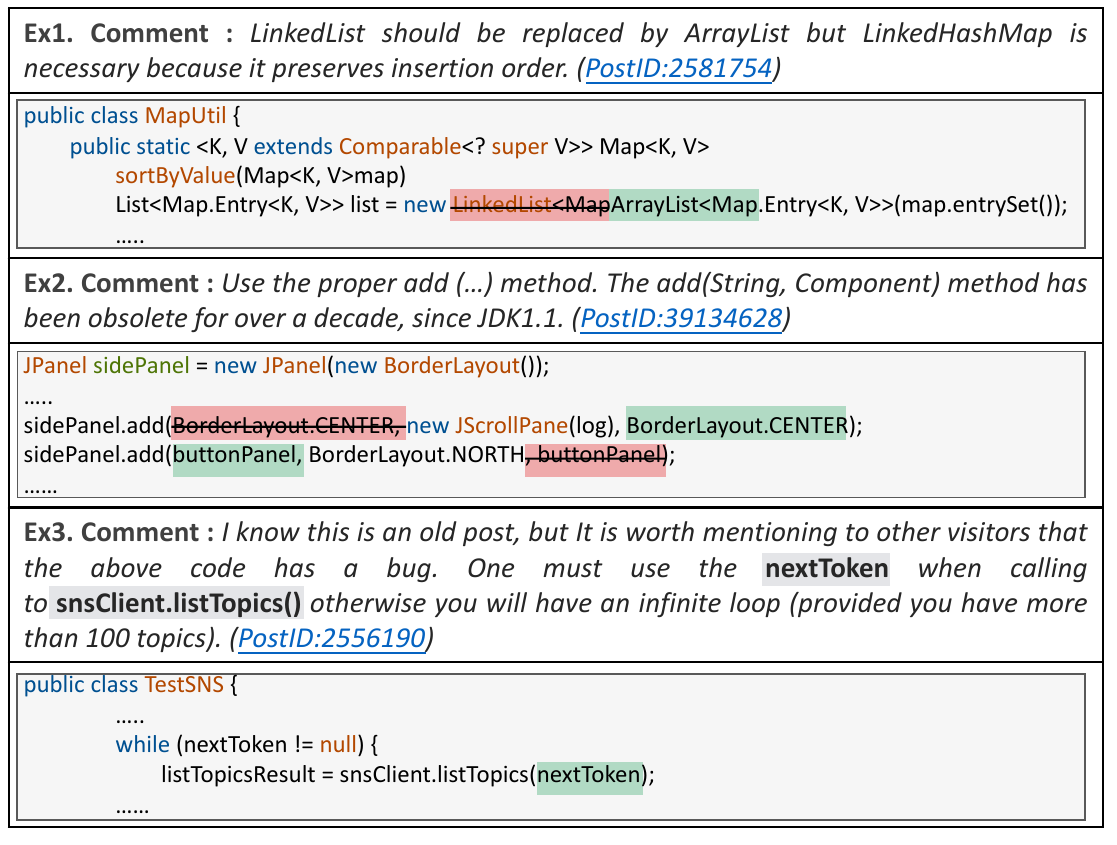}
\caption{Motivating Examples}
\vspace{-0.3cm}
\label{fig:moti}
\end{figure}

Although SO code snippets may not always be flawless, the platform's comments serve as valuable resources for identifying potential issues and offering constructive feedback.
Fig.~\ref{fig:moti} shows several comment-edit examples in SO to deliver helpful information and knowledge for different problems. 
(1) Even popular SO answers may contain problematic code snippets.
For example, Ex.1 presents a suboptimal code snippet, the question post ``\textit{sort a Map$\langle$Key, Value$\rangle$ by values}'' received more than 1.8k votes and has been visited for more than 1.8m times since its creation, and this code snippet received more than 1k votes by SO users. 
However, as suggested by a developer, this code snippet is suboptimal by using \texttt{LinkedList} instead of \texttt{ArrayList}. 
After the comment was posted, the answer post was subsequently updated accordingly. 
However, it is worth noting that this problematic code had been present on SO for over eight years before any editing took place. 
(2) SO answers may become obsolete or outdated as a result of the rapid evolution of software systems.
Ex.2 demonstrates an example where the code snippet used an outdated method \texttt{add} which has been obsolete for over a decade. 
(3) SO answers may contain buggy code. 
An example is shown in Ex.3, as the commenter pointed out, ``\textit{I know this is an old post, but it is worth mentioning to other visitors the above code has a bug. One must use \texttt{nextToken} ..., otherwise you will have an infinite loop}''. 
This answer post was created in August 2014, but this bug has not been discovered until July 2015.  
During this time, this code snippet was visited by SO users thousands of times, which can mislead developers and cause the introduction of bugs in their subsequent development. 

SO comments provide valuable information to improve the code snippet's quality, however, prior studies~\cite{ragkhitwetsagul2019toxic, zhang2019empirical} found that answers are rarely updated after comments are posted, and only 4.6\% of the answers are edited after any comments~\cite{soni2019analyzing}. 
In other words, the valuable information hidden in comments  is mostly ignored and the comments are rarely integrated back into answers.
\zp{
Therefore, in this work, we aim to fill this gap by proposing a tool to automatically update code snippets based on their associated comments, which can enhance the answer quality and reduce the likelihood of introducing bugs.
}

\section{Preliminary Investigation}
\label{sec:pre}

\begin{figure}[]
\centering
\includegraphics[width = 0.99\linewidth]{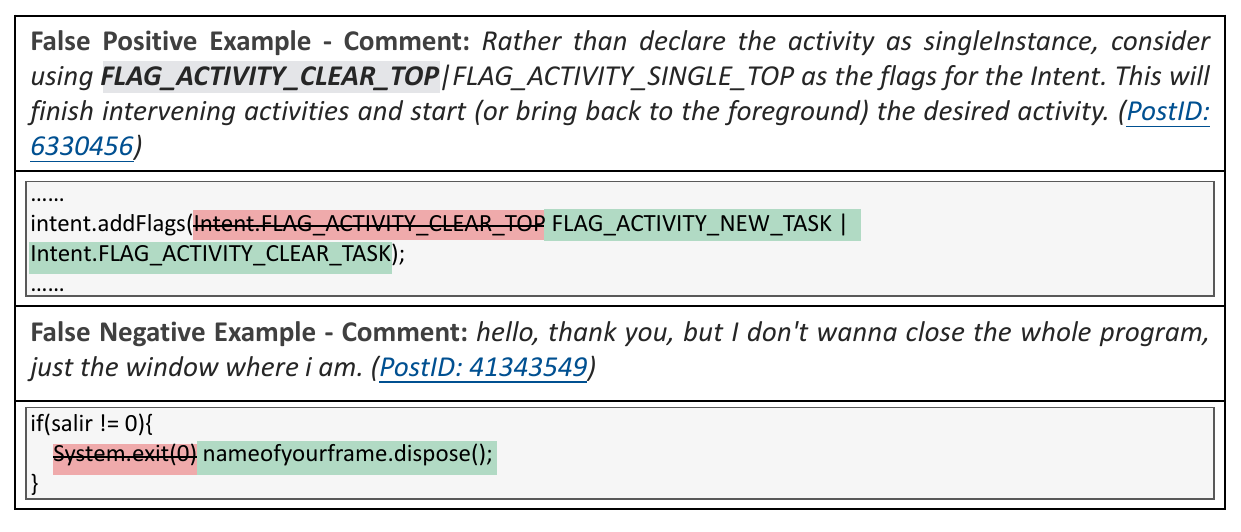}
\caption{Preliminary Investigation Examples}
\vspace{-0.3cm}
\label{fig:pre}
\end{figure}

Tang et al.~\cite{tang2021using} conducted the first study on how to map a comment with its associated edit. 
They proposed a simple matching-based method to determine whether an edit is related to a comment.
Specifically, their method matches a comment to an edit based on three heuristic rules: (1) the comment occurred before the edit; (2) the comment mentions a code term that gets added or removed from a code snippet within the edit; (3) the commenter and the editor are different users. 
To evaluate their method performance, Tang et al. manually annotated 194 comment-edit pairs (from 100 SO posts) to create a ground truth dataset for estimating the precision and recall of their approach.  
As they reported, their matching-based method achieved 70\% precision and 32\% recall on this ground truth dataset. 
They further analyzed 1,910 positive cases (including 382 Java cases) predicted by their method and reported a precision of 78\%.  
To gain a deeper understanding of their method and dataset, we manually investigated 382 Java comment-edit pairs predicted as relevant by their method and found the following limitations:
\begin{itemize} 
    
    \item Tang's method of using human-defined rules to predict the relevance of edits and comments has two disadvantages: first, it \textbf{ignores the semantic relationship between code terms and edits}; second, \textbf{relevant comment-edit pairs are not necessarily valid comment-edit pairs}.
    Fig.~\ref{fig:pre} shows a false positive example, the comment mentions the code term \texttt{FLAG\_ACTIVITY\_CLEAR\_TOP} that was deleted in the edits, Tang's method predicted this comment-edit pair as relevant.  
    However, the edit did not modify according to the comment's suggestions (use \texttt{FLAG\_ACTIVITY\_SINGLE\_TOP}). 
    Such pairs are considered as relevant by Tang et al. but failed to meet our valid comment-edit pair definition. 
    We relabeled these 382 pairs with standards of our valid comment-edit pairs and reevaluated Tang's method, the precision of Tang's method significantly dropped from 70\% to 56\%. 
    
    \item \textbf{There are a large number of SO comments that caused an edit but did not contain any code terms,} resulting in a large number of false negative samples.  
    Fig.~\ref{fig:pre} shows a false negative example where the comment doesn't mention any code terms. 
    Tang's method can't identify such comment-edit pairs because the comment does not use explicit code terms but rather explains the problem and how the code can be fixed. 
    This limitation makes their method miss a significant number of valid comment-edit pairs, resulting in notably low recall scores. 
\end{itemize}

Although Tang et al. constructed the first relevant comment-edit pair dataset (including 248,339 pairs), \textbf{the proportion of valid comment-edit pairs is only 56\%. 
In other words, nearly half of their dataset samples are noise, the low quality of their dataset poses substantial threats and risks for researchers or studies intending to reuse it. 
} 
Therefore, it is necessary to build a high-quality dataset of valid comment-edit pairs.

\section{Approach}
\label{sec:approach}
\begin{figure} 
	\centering
	\includegraphics[width = 1.0\linewidth]{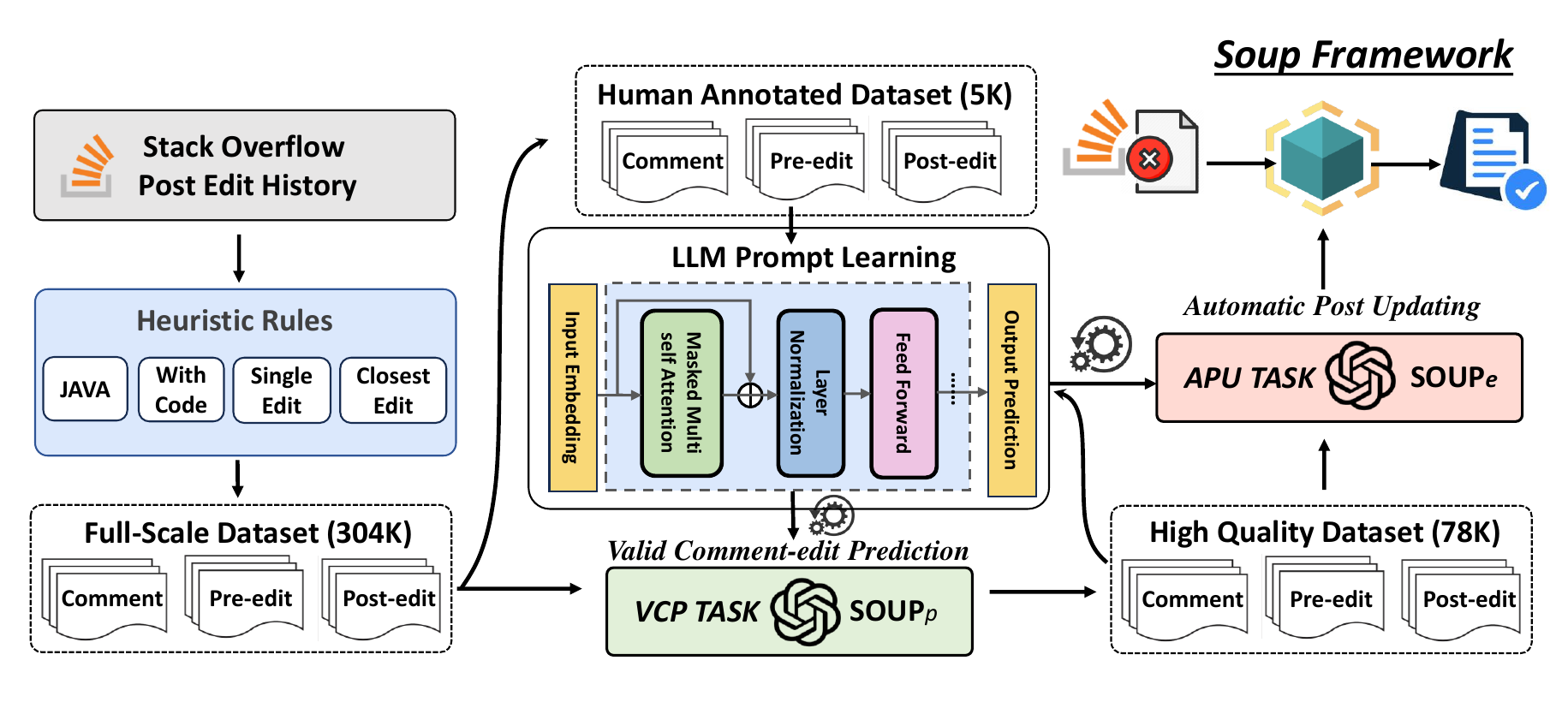}
	\caption{Workflow of Our Approach}
	\label{fig:workflow}
\end{figure}

In this section, we first formally define our two key tasks, i.e., valid comment-edit prediction (referred to as VCP) and automatic post updating (referred to as APU). We then describe the details of our approach to solving these two tasks. The workflow of our approach is shown in Fig.~\ref{fig:workflow}.


\subsection{Task Definition}

\subsubsection{Valid Comment-edit Prediction}
\zp{
On SO, an edit $\mathbf{e}$ is associated with a pair of posts before and after editing, denoted as $\mathbf{pre_{e}}$ and $\mathbf{post_e}$. 
Formally, let $\mathbf{c}$ be the candidate comment to be checked, the VCP task is to find a function \texttt{Predict} so that:  
\begin{equation}
\label{eq:Predict}
\texttt{Predict} ( \mathbf{pre_{e}}, \mathbf{post_{e}}, \mathbf{c} ) =   
\left\{ \begin{array}{rcl}
1 & \mbox{for} & \mathbf{triplet} \in \mathbf{v} \\
0 & \mbox{for} & otherwise
\end{array}\right. 
\end{equation}
where $\mathbf{triplet} \in \mathbf{v}$ denotes that the triplet ($\mathbf{pre_{e}}, \mathbf{post_{e}}, \mathbf{c}$ ) is valid (according to the definition in the introduction). 
}

\subsubsection{Automatic Post Updating}
\zp{The APU task is to update code snippets in SO posts based on their associated comments.
More formally, it aims to find a function  so that: 
\begin{equation}
\label{eq:update}
\texttt{Update} (\mathbf{pre_{e}}, \mathbf{c}) = \mathbf{post_{e}}
\end{equation}
given the post's pre-editing version $\mathbf{pre_{e}}$ and its associated comment $\mathbf{c}$ as input, $\texttt{Update}$ aims to generate a post's post-editing version $\mathbf{post_{e}}$ as output. 
\textsc{Soup} framework leverages the LLM-based model to approximate \texttt{Predict} and \texttt{Update} respectively.
}


\subsection{Valid Comment-edit Prediction}
The goal of the VCP task is to obtain a high quality dataset of valid comment-edit pairs. 
\zp{
We translate this goal into a valid comment-edit prediction task. 
For this task, we first collected a full-scale dataset for potential comment-edit pairs by applying a set of pre-defined rules. 
After that, we sampled 5,000 comment-edit pairs from the full-scale dataset to perform manual annotation.  Such a sample guarantees a margin of error of ±2\% with a confidence level of 99\%~\cite{rosner2006fundamentals}.
Lastly, we used 80\% of the 5,000 annotated comment-edit pairs to fine-tune LLMs (CodeLlama in this study) for this VCP task, adapting the LLM to effectively distinguish valid comment-edit pairs and invalid ones. 
The approach details are as follows:
} 
\subsubsection{\label{sec:data_all}Full-scale dataset preparation.} 
\zp{To perform VCP task, we first make a full-scale dataset for comment-edit pairs by mining edit histories of SO posts. 
To make a fair comparison, we use the same version of SOTorrent~\cite{baltes2018sotorrent} dataset (September 23, 2019) as Tang et al.~\cite{tang2021using}, which captures the edit history of all SO posts. 
Following Tang's research, we fetch all comment-edit pairs.
Subsequently, we make our full-scale dataset by filtering with the following criteria: 
\begin{itemize}
    \item We exclusively focus on SO Java posts in this study, since Java is one of the most widely used programming languages. It is worth mentioning that our work can be readily extended to other programming languages.  
    \item Our goal is to update code snippets based on their corresponding comments. Therefore, we only consider answer posts with code snippets. 
    \item An answer post can be associated with multiple code snippet blocks, in this study, we only consider answer posts that contain edits to a single existing code block. 
    Edits to multiple code blocks at the same time are excluded. 
    \item The edit must occur after the comment, and if multiple edits occur after a given comment, we paired the comment with its ``closest'' edit. 
\end{itemize}
It is worth mentioning that we didn't consider the second rule (i.e., a comment mentions a code term that gets removed or added in a later edit) and the third rule (i.e., the commenter and editor are different users) proposed by Tang et al~\cite{tang2021using}, because we want to make a full-scale dataset to include as many potential comment-edit pairs as possible, and covering wider usage scenarios (comments without code terms and their associated edits). 
Based on the above criteria, we filtered out 304,567 potential Java comment-edit pairs as our full-scale dataset.   
}

\subsubsection{Human annotation}
\zp{After collecting the full-scale dataset of 304,567 SO comment-edit pairs for Java, we sampled 5,000 pairs for manual annotation. 
Considering the limitation of our hardware resources, we set the maximum sequence length of the comment-edit pairs to 2,048 tokens in this study. 
Since we focus on updating code snippets, we removed code comments within the code snippets. 
Then two authors are asked to perform the manual annotation on the above 5,000 comment-edit pairs independently. 
These two annotators are software engineering researchers with more than 10 years of Java programming language and have used Stack Overflow for more than 7 years. 
For each comment on an answer, the two annotators separately analyze the comment-edit pair to determine if it is valid based on the definition provided in the introduction. 
The total time spent by the two annotators to complete the manual annotation costs 127.7 hours and 177.2 hours respectively, with an average of 1min 32s and 2min 8s for labeling one comment-edit pair. The Cohen's kappa coefficient~\cite{cohen1968weighted} for the two annotators is 0.795, indicating substantial agreement between them~\cite{landis1977measurement}.
}

\zp{After each annotator performs the above labeling process independently, there are 8.4\% instances of disagreements. 
In this study, we resolve the disagreements as follows: when different opinions are met, the first author will act as a mediator to discuss with two annotators, then make his/her own decision on this conflict case. 
The final label will be determined by majority voting. 
Finally, 1356 out of 5000 comment-edit pairs are labeled as valid, while the remaining 3644 edit-comment pairs are labeled as invalid. 
Notably, the label distribution of the dataset is slightly imbalanced, according to the research of Krawczyk et al.~\cite{krawczyk2016learning}, the slight imbalance ratio has a relatively small impact on the model performance. 
In addition, considering that our dataset reflects the real-world label distributions, we directly use this slightly imbalanced dataset to train and evaluate the model.}

\zp{After constructing the human annotated dataset, we randomly split the dataset into train/validation/test set by 8:1:1, the train set is used to fine-tune LLMs for the VCP task, the validation set is used to minimize the overfitting, and the test set is only used for evaluating the model performance on VCP.
The details of train, validation and test set of the human annotated dataset are displayed in the first two rows of Table~\ref{tab:data_vcp}. 
}

\begin{table}
\footnotesize
  \caption{Dataset Information}
  \label{tab:data_vcp}
  \centering
  \begin{tabular}{lcccc}
    \toprule
    \textbf{Dataset} & \textbf{Label} & \textbf{Train} & \textbf{Validation} & \textbf{Test} \\
    \midrule
    \multirow{2}{*}{Human annotated} & Valid & 1094 & 135 & 127 \\
    & Invalid & 2906 & 365 & 373 \\
    \midrule
    High-quality & / & 62,664 & 7,827 & 7,826 \\
  \bottomrule
\end{tabular}
\end{table}

\subsubsection{VCP Modeling}
\zp{The manual annotation process is time-consuming and label-intensive, to create a high quality dataset, we need to design effective models to perform valid comment-edit prediction automatically. 
Inspired by the great potential of large language models (LLMs) in code comprehension, we fine-tune a LLM to perform the VCP task with prompt learning by using our human annotated dataset. 
We refer our model on the VCP task as \textsc{Soup}$_p$. 
Recently, prompt learning has been widely used with LLMs to support various tasks~\cite{lester2021power, yang2024automatic, huang2022prompt}. 
Prompt learning combines the advantage of prompt engineering and fine-tuning. 
It utilizes the prompt to elicit the domain knowledge of LLM and fine-tuning to adapt their learned representations to specific downstream tasks. 
In order to design an effective prompt for the VCP task,  we followed the prompt engineering guidelines mentioned in the previous studies~\cite{feng2024prompting, wei2022chain} to construct and iteratively optimize the prompt based on the validation set, which included: (1) writing specific judgment criteria based on the task definition and the annotation experience of two annotators; (2) asking for a structural output (e.g., ``\#\#\#Response: \textless label\textgreater\textless/s\textgreater''); (3) using delimiters to clearly indicate distinct parts (e.g., use \textless\textgreater~to wrap the code); (4) analyzing the error cases on the validation set to iteratively optimize the prompt. 
We detailed the description of our prompt for judging valid comment-edit pairs in Table~\ref{tab:prompt_vcp}, the complete prompt can be found in our replication package~\cite{replicate}.
}

Regarding LLMs, we chose CodeLlama as the base LLM in our research. 
CodeLlama is a decoder-only transformer model, created by further training Llama2~\cite{touvron2023llama} on code-specific datasets. As one of the most popular open-source code LLMs, it supports multiple programming languages including Java and has shown impressive capabilities in code-related tasks. 
Considering the computation resource limitation, we chose the CodeLlama instruct model with 13B parameters~\cite{roziere2023code} in this study and fine-tuned it using the low-rank adaptation (LoRA) method~\cite{hu2021lora}.

\begin{table}[h]
\footnotesize
\renewcommand{\arraystretch}{1}
\setlength{\tabcolsep}{5pt} 
\caption{Prompt for Valid Comment-edit Prediction}
\label{tab:prompt_vcp}
\centering
\begin{tabularx}{\columnwidth}{m{0.1\columnwidth}X}
\toprule
\textbf{Prompt} & \textbf{Instantiation} \\
\hline
Input & \textless s\textgreater  
~Please check if the code\_before, comment, and code\_after simultaneously meet the following four conditions. \newline
Condition 1: The comment points out a general or specific flaw in code\_before, or provides a general or specific suggestion for code\_before, or raises a general or specific question about code\_before. \newline
Condition 2: code\_after has optimized all the flaws pointed out in the comment, implemented all the suggestions given in the comment, and resolved all the questions raised in the comment.\newline
Condition 3: code\_after has not made any changes beyond those mentioned in the comment.\newline
Condition 4: code\_after has not introduced any new errors.\newline
Please answer with yes or no whether they meet the aforementioned four conditions simultaneously. \\
\hline
Output & \#\#\#Response: \textless label\textgreater\textless/s\textgreater \\
\bottomrule
\end{tabularx}
\end{table}


\begin{table}[]
\footnotesize
\renewcommand{\arraystretch}{1}
\setlength{\tabcolsep}{5pt} 
\caption{Prompt for Automatic Post Updating}
\label{tab:prompt_apu}
\centering
\begin{tabularx}{\columnwidth}{m{0.1\columnwidth}X}
\toprule
\textbf{Prompt} & \textbf{Instantiation} \\
\hline
Input & \textless s\textgreater 
~I want you to act as a code optimizer, you need only to return the optimized code snippet to me. Below is an instruction that describes a task. Write a response that appropriately completes the request.\newline
Code snippet: \textless\textgreater\newline
Comment: \textless\textgreater\newline
\#\#\#Input: Please optimize the code snippet based on the comment.\\
\hline
Output & \#\#\#Response: \textless code\_after \textgreater\textless/s\textgreater \\
\bottomrule
\end{tabularx}
\end{table}

\subsection{Automatic Post Updating}
\subsubsection{High quality dataset construction}

The \textsc{Soup}$_p$ model is trained to label valid comment-edit pairs automatically and effectively. 
According to our evaluation results, \textsc{Soup}$_p$ has achieved a precision of 80.8\% and a recall of 74.0\% on the VCP task (details in Section~\ref{sec:eval}). 
Considering that there is also an 8.4\% inconsistency in the manual annotation process, we consider \textsc{Soup}$_p$ as an effective model for identifying valid comment-edit pairs, and has achieved a human-like performance. 
Therefore, given its remarkable performance of \textsc{Soup}$_p$ for predicting valid comment-edit pairs, we apply \textsc{Soup}$_p$ back to our full-scale dataset (304K comment-edit pairs) to automatically identify valid pairs among them and construct a large-scale high quality dataset. 
Particularly, for a given potential comment-edit pair from the full-scale dataset, we use \textsc{Soup}$_p$ to automatically label it as valid or invalid. 
Finally, we obtain 78,317 valid comment-edit pairs as our large-scale high quality dataset. 
\textsc{Soup}$_p$ achieves a precision of 80.8\%, which means that there is 19.2\% noise data in this dataset. 
To reduce the noise data ratio, one possible way is to conduct manual verification according to our method of manual annotation, but it is too time-consuming and labor-intensive. 
Another way is to craft manually defined rules for different types of noisy data, but it requires expert knowledge and additional annotation efforts. 
We plan to design automated methods that do not rely on manual annotations or rules in the future to further improve the model's precision and reduce the noise ratio of the dataset.


\zp{After constructing the large-scale high quality dataset, we randomly split the dataset into train/validation/test set by the ratio of 8:1:1. 
The train set is used to fine-tune LLMs on the APU task, the validation set is used to assess how well the model generalizes to new samples. 
The test set is used to estimate the model's performance on unseen data. 
The details of the train, validation and test set of the high quality dataset are demonstrated in the last row of Table~\ref{tab:data_vcp}.
}


\subsubsection{APU Modeling}
\zp{Similar to VCP modeling, for APU, we fine-tuned CodeLlama with prompt learning on the high-quality dataset built by \textsc{Soup}$_p$ using LoRA.
The only difference is the input prompts, the prompts designed for APU task is shown in Table~\ref{tab:prompt_apu}. 
We refer to our model on the APU task as \textsc{Soup}$_u$. 
It is worth mentioning that our framework is not restricted to CodeLlama, CodeLlama can be easily replaced by other LLMs or even pre-trained language models (PLMs) according to different users' computational resources or preferences. 
}

\subsection{\textbf{Implementation Details}}
In this work, we use the Hugging Face library for implementation and fine-tuning. The fine-tuning was conducted on a server equipped with four NVIDIA A800 80GB GPUs. 
The fine-tuning process follows these hyperparameters settings: learning rate of 2e-4, batch size of 16, for a total of 5 epochs. The models with the best performance on the validation set are selected for final evaluation. 

\section{Evaluation}
\label{sec:eval}
In this section, we evaluate how effective is our proposed framework \textsc{Soup} for VCP and APU tasks. 
We aim to answer the following four key research questions:

\begin{itemize}
    \item \textbf{RQ1:} How effective is our approach for predicting valid comment-edit pairs?
    \item \textbf{RQ2:} How effective is our approach for automatically updating SO posts based on their comments?
    \item \textbf{RQ3:} To what extent does the dataset influence model performance in the APU task?
    \item \textbf{RQ4:} How acceptable are our updated SO posts in real-world scenarios?
\end{itemize}

\subsection{RQ1: Effectiveness of \textsc{Soup} on VCP} 
\subsubsection{Experimental Setup}
Regarding the VCP task, we fine-tuned CodeLlama with 80\% of our human annotated dataset (i.e., train set). 
Then the model with best performance on the validation set is chosen as \textsc{Soup}$_p$ for this VCP task, and the test set is used for evaluating our model and baselines. 


\subsubsection{Evaluation Metrics and Baselines}


The VCP task can be regarded as a binary classification task, we use \textit{Precision}, \textit{Recall} and \textit{F1-score} as evaluation metrics. These metrics effectively gauge both the accuracy and comprehensiveness of the resulting APU dataset. We use the manually curated test set from Table~\ref{tab:data_vcp} to evaluate the performance of different models, which are as follows:
\begin{itemize}
    \item \textbf{Tang's:} Tang et al.~\cite{tang2021using} first proposed a matching-based approach for related comment-edit prediction, in particular, their method determines whether a comment is related to an edit as follows: if the comment mentions a code term that gets added or deleted in the comment, then this pair is predicted as related. We run Tang's method on our human annotated test set for evaluation. 
    \item \textbf{\textsc{Soup}$_p$:} CodeLlama trained on the training set from Table~\ref{tab:data_vcp}, which achieved the best performance on the validation set.
\end{itemize}

\subsubsection{Experimental Results}
\zp{The experimental results for RQ1 is shown in Table~\ref{tab:RQ1_results}. 
From the experimental results, we can see that:
(1) Tang's method achieved a recall of 12.6\%, the reason is that their method was unable to identify valid comment-edit pairs, in which comment did not involve code element additions or deletions. 
As shown in Ex1 of Fig.~\ref{fig:rq1}, the comment doesn't even contain any code elements. 
The extremely low recall rate indicates that Tang's method misses a large number of valid pairs. 
(2) Tang's method achieved a precision of 57.1\%, which is consistent with the 56\% precision in our preliminary experiments. 
The reason is that the match-based method cannot capture the semantics between comments and edits, making it difficult to determine whether they are valid comment-edit pairs. 
As shown in Ex2 of Fig.~\ref{fig:rq1}, Tang's method wrongly predicted it as a positive case. 
Although the updated code deleted the code element "\texttt{new}" mentioned in the comment, it had nothing to do with comment suggestions (e.g., ``\textit{the builder must build a client}''). 
In other words, this is an invalid comment-edit pair. 
Our \textsc{Soup}$_p$ labeled this pair as negative (i.e., invalid) by successfully capturing the semantic relationship between the comment and edit. 
(3) Our method (i.e., \textsc{Soup}$_{p}$) outperforms Tang's method in terms of all evaluation metrics. 
We attribute this to the following reasons: firstly, compared with Tang's method, our approach is based on LLM, which shows great potential for code comprehension. 
In other words, LLM has its own ability to capture the semantic relationship between the comment-edit pairs. 
Secondly, our well designed prompts further elicit the knowledge from CodeLlama and adapt it to our VCP task.
}

\begin{table}[h]
\footnotesize
  \caption{Results of Different Methods on VCP}
  \label{tab:RQ1_results}
  \centering
  \begin{tabular}{lccc}
    \toprule
    \textbf{Approach} & \textbf{\textit{Precision}} & \textbf{\textit{Recall}} & \textbf{\textit{F1-score}}\\
    \midrule
    Tang's  & 57.1\% & 12.6\% & 20.6\%\\
    \midrule
    \textsc{Soup}$_p$ & \textbf{80.8\%} & \textbf{74.0\%} & \textbf{77.3\%} \\
  \bottomrule
\end{tabular}
\end{table}

\subsubsection{Manual Analysis}

\begin{figure}[]
\centering
\includegraphics[width = 0.99\linewidth]{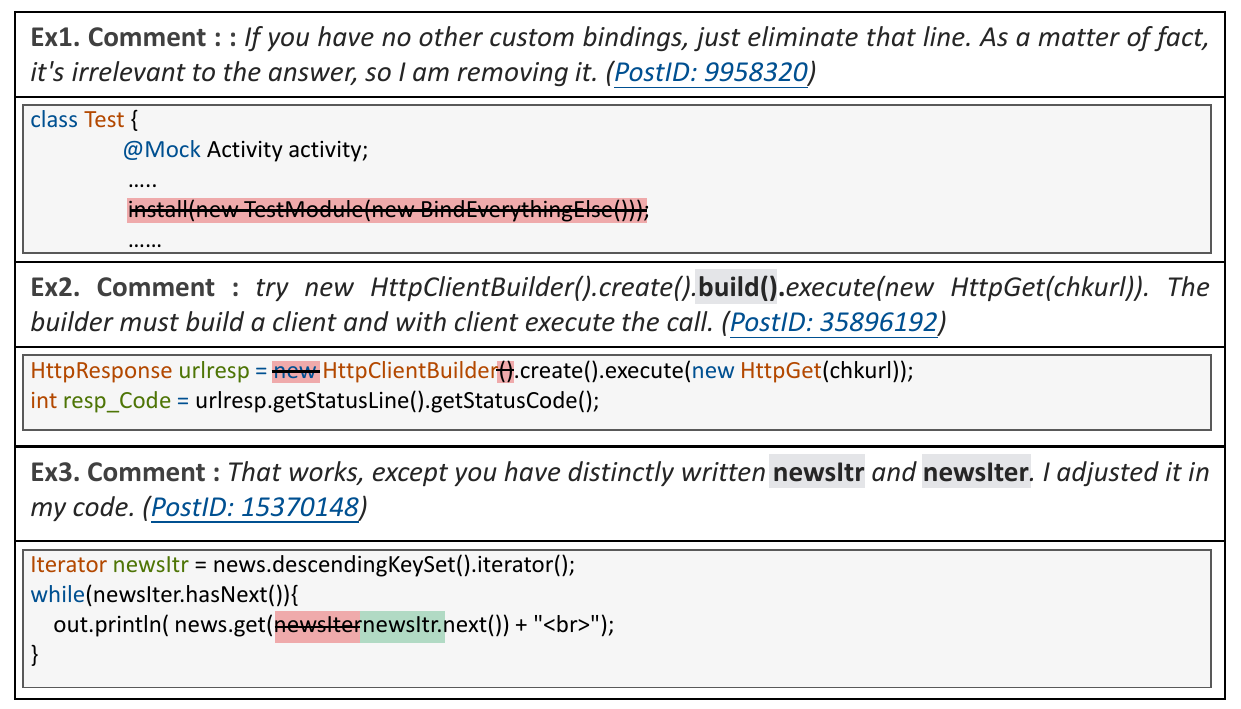}
\caption{Examples of Errors in Different Methods}
\vspace{-10pt}
\label{fig:rq1}
\end{figure}

We manually analyzed all the false positive cases (23 cases) and false negative cases (30 cases) predicted by our approach in RQ1. 
The false positive cases are mainly caused by the following reasons: 
(1) The comment raised multiple issues or suggestions, and the edits only addressed part of them (10 cases). 
Fig.~\ref{fig:rq1} Ex.3 demonstrates such a case where the comment points out the inconsistency between the variable names \textit{newsItr} and \textit{newsIter}. The updated code only modifies one instance of \textit{newsIter}, but the issue still persists in the \textit{while} statement. Therefore, it does not meet the definition of a valid comment-edit pair (as it does not fully address the issue raised in the comment) and should be considered a negative example. The model may consider such comments helpful for code changes, thus classifying them as positive cases. (2) Comments are not beneficial to code changes (13 cases). 
These comments usually contain some actions related to editing (e.g., "\textit{I will edit it later}"), which mislead the model into thinking that they have contributed to the editing, thus determining them as positive cases. 

The false negative cases are mainly caused by the following reasons: (1) Code changes occur within 5 tokens (22 cases). 
Minor changes may not provide sufficient context for the model to infer, thus judging comments as irrelevant to the edits. 
(2) The functionality of the code has undergone significant changes after editing (8 cases). 
The comments of such edits are usually new functional requirements. These code changes or updates are too complicated for our model to perfectly handle, thus wrongly determining them as negative cases. 
To alleviate the above issues, a possible optimization direction is to utilize data augmentation techniques to further enhance LLM’s code comprehension capabilities on such data samples. 

\find{
\textbf{Answer to RQ1: How effective is our approach for predicting valid comment-edit pairs?}
\textsc{Soup}$_{p}$ achieved an accuracy of 80.8\% and a recall rate of 74.0\% on the VCP task, far surpassing the current state-of-the-art method.
}


\subsection{RQ2: Effectiveness of \textsc{Soup} on APU}

\subsubsection{Experimental Setup}
For the automatic post updating task, we fine-tuned CodeLlama with the train set of the high quality dataset, the model with best performance on the validation set is chosen as \textsc{Soup}$_u$ for this APU task. 
Then the test set is used for estimating the performance of our approach and other baselines.

\begin{table}
\footnotesize
  \caption{Results of Different Methods on APU}
  \label{tab:rq2_result}
  \centering
  \begin{tabular}{lcc}
    \toprule
    \textbf{Approach} & \textbf{EM} & \textbf{CB} \\
    \midrule
    CodeLlama$_{pr}$ & 0.2\% & 33.8\% \\
    \midrule
    ChatGPT  & 2.1\% & 59.6\% \\
    \midrule
    \textsc{Soup}-CodeLlama & 25.6\% & 73.5\% \\
    \midrule
    \textsc{Soup}-CoditT5 & \textbf{25.9}\% & \textbf{76.4\%} \\
  \bottomrule
\end{tabular}
\end{table}

\subsubsection{Evaluation Metrics and Baselines}
\zp{The APU task aims to generate the correct updating code for SO posts, which can be regarded as a generation task. 
Therefore, we choose the widely used Exact Match (\textit{EM})  and CodeBLEU (\textit{CB})~\cite{ren2020codebleu} as evaluation metrics for this RQ. 
\textit{EM} metric is used to evaluate whether the model prediction exactly matches the ground truth. 
\textit{CB} is a common metric that measures lexical overlap for text (code) generations. 
The following baselines are employed: 
\begin{itemize}
    \item \textbf{\textsc{Soup}-CodeLlama}: 
    We fine-tuned CodeLlama with prompt learning on the training set of the high quality dataset, this model is denoted as \textbf{\textsc{Soup}-CodeLlama}. 
    \item \textbf{\textsc{Soup}-CoditT5}: 
    Zhang et al.~\cite{zhang2022coditt5} fine-tuned a model CoditT5 for software-related editing tasks using the pre-tained language model CodeT5~\cite{wang2021codet5}. 
    Their approach showed promising performance 
    on various downstream tasks (e.g., comment updating~\cite{panthaplackel2020learning}, bug fixing~\cite{tufano2019learning} and automated code review\cite{tufano2022using}). 
    The APU task is a similar task related to software editing, the difference is our task aims to update SO code snippets based on their comments. 
    We further fine-tuned CoditT5 on the APU task with our \textsc{Soup} framework, denoted as \textbf{\textsc{Soup}-CoditT5}. 
    \item \textbf{ChatGPT:} ChatGPT has demonstrated powerful ability on code generation~\cite{clement2020pymt5,feng2020codebert,fried2022incoder}, test generation~\cite{chen2022codet} or bug fixing~\cite{sobania2023analysis}. 
    Applying prompt engineering to guide ChatGPT to complete particular tasks was widely explored by different researchers. 
    In our work, we adopted ChatGPT based on GPT-3.5-turbo as a baseline and evaluated ChatGPT's performance on the APU task by using the similar prompt as shown in Table~\ref{tab:prompt_apu}.
    \item \textbf{CodeLlama$_{pr}$}: 
    In order to explore the effectiveness of employing the fine-tuning strategy, we directly tested the performance of CodeLlama on the APU task without extra fine-tuning, only using the prompts from Table~\ref{tab:prompt_apu}.
\end{itemize}
}

\subsubsection{Experimental Results}
The experimental results of RQ2 are demonstrated in Table~\ref{tab:rq2_result}. 
From the table, we observe the following points: 
(1) Directly using the prompt engineering approach with LLMs is not suitable for our APU task. 
For example, CodeLlama$_{pr}$ (13B) achieved the worst performance, and even using ChatGPT (175B) only reached a 2.1\% exact match, indicating that relying solely on prompts is insufficient to complete the APU task. 
(2) After fine-tuning with our high quality dataset, CodeLlama's \textit{EM} value sharply increased from 0.2\% to 25.6\% (i.e., the result of \textsc{Soup}-CodeLlama). 
This demonstrates that fine-tuning LLMs with large-scale high quality dataset can effectively adapt LLM's learned feature representation to specific tasks, significantly enhancing their performance on the APU task.
(3) Fine-tuning LLMs/PLMs with our \textsc{Soup} framework achieve best performance on this task. 
After fine-tuning, \textsc{Soup}-CoditT5 (with only 220M parameters) can achieve a comparable or even better performance than \textsc{Soup}-CodeLlama. 
Our \textsc{Soup} framework is highly flexible and can be integrated with any open-source LLMs or PLMs, benefiting users to choose according to their own hardware resources and preferences. 

\subsubsection{Manual Analysis}

\begin{table}
\footnotesize
  \caption{Comment Categories for Manually analyzed examples}
  \label{tab:rq2_error}
  \centering
  \begin{tabular}{lcc}
    \toprule
    \textbf{Category} & \textbf{Positive} & \textbf{Negative} \\
    \midrule
    Improvement  & 28 & 5 \\
    \midrule
    Weakness & 14 & 27  \\
    \midrule
    Inquiry  & 7 & 4  \\
    \midrule
    Addition  & 1 & 2 \\
    \midrule
    Total  & 50 & 38 \\
  \bottomrule
\end{tabular}
\end{table}

\begin{figure}[]
\centering
\includegraphics[width = 0.99\linewidth]{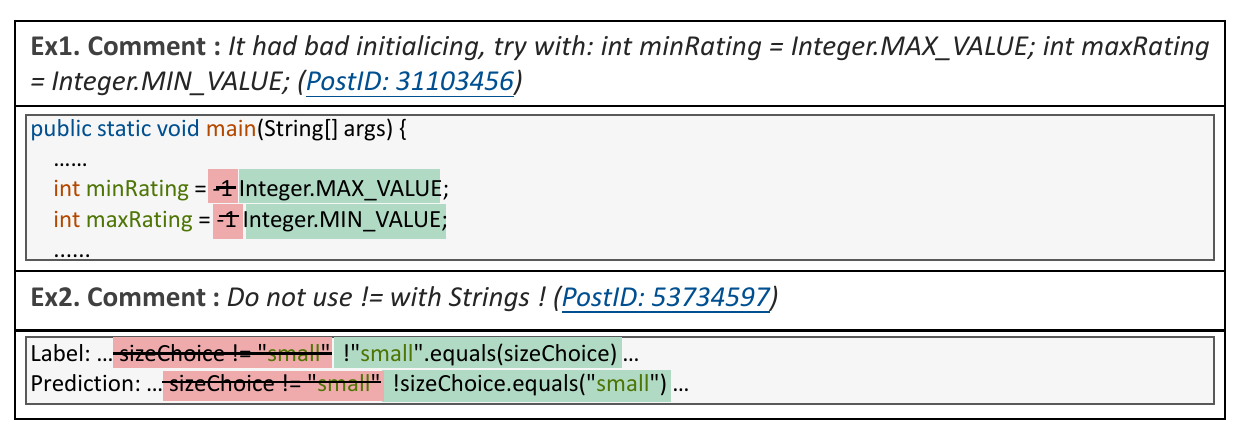}
\caption{Positive and Negative Examples of Soup on APU}
\vspace{-0.3cm}
\label{fig:rq2_exam}
\end{figure}

We sampled from the results of \textsc{Soup}-CodeLlama for manual analysis (considering \textsc{Soup}-CodeLlama is more robust to noisy data and generalizes better on new datasets in RQ3). 
Considering the same comment can be addressed with different ``correct'' updatings, we manually examined 50 exact matching cases (denoted as positive) and 50 non-exact matching cases (denoted as negative) for analysis. 
The positive cases are samples with \textit{CodeBLEU} equal to 1, while the negative cases are samples with \textit{CodeBLEU} score less than 1. 
Specifically, we sampled 25 negative samples with \textit{CodeBLEU} less than 0.5 and the other 25 negative samples with \textit{CodeBLEU} larger than 0.5. 
We carefully investigate these samples to see if their edits address the comment or not. 
After our manual analysis, 12 non-exact matching samples were incorrectly labeled by \textsc{Soup}$_p$ in the data preparation stage, these are invalid comment-edit pairs that should be removed from training or testing. 
Only 38 non-exact matching samples were verified as valid comment-edit pairs, this is close to the precision (80.8\%) we found in RQ1. 
Furthermore, we categorized these positive and negative examples according to the primary comment categories defined by Zhang et al.~\cite{zhang2019reading}, and the results are shown in Table~\ref{tab:rq2_error}.  

From the table, it can be seen that: 
(1) {Soup}$_{u}$ is good at updating SO posts with improvement comment (28 cases). 
This is because improvement comments typically contain specific suggestions for changes, which the model can easily understand the intention to modify. 
Fig.~\ref{fig:rq2_exam} Ex1 shows one such case, where the comment explicitly suggested the use of statements ``\texttt{int minRating = Integer.MAX\_VALUE;}'', and the according generated edits fully implements this suggestion. 
(2) {Soup}$_{u}$ is not so good at handling post updates with comments categorized as weakness. 
Regarding the weakness comment type, 17 comments do not agree with the original code's functionality (addressed his/her concerns), such as the comment ``\textit{Above solutions doesn't work if the list is empty, but I still need to handle that case}~\cite{rq2_so}''. 
The edits corresponding to these comments usually implement new features, and the code undergoes significant changes before and after the update. The model must infer the actual functionality requests of the commenter and modify the original code thoroughly, or even regenerate code from scratch.   
These edits are too complicated for {Soup}$_{u}$ to perfectly handle. 
A possible solution is to add more contextual information for these updates, such as the question post. 
It would be interesting to address these limitations, but it is beyond the scope of this work. 
(3) There are 7 negative cases that fully address their comments equivalently but not with exact matches. 
Ex2 in Fig~\ref{fig:rq2_exam} demonstrates such a case. 
The \textit{CodeBLEU} between our generated edit and the ground truth edit is only 0.32. 
The ground truth edit is \texttt{!"small".equals(sizeChoice)} while the edit generated by our approach is \texttt{!sizeChoice.equals("small")}, these two edits are equivalent and both can fully address its comment. 
This illustrates the \textit{EM} metric just estimates the lower bound of our model, if we include these equivalent samples, the model performance can be further improved. 

\find{\textbf{Answer to RQ2: How effective is our approach for automatically updating SO posts based on their comments?}
\textsc{Soup}$_{u}$ achieved a 25.6\% exact match rate, particularly excelling in solving improvement-type updates.
}

\subsection{RQ3: Cross Dataset Evaluation}
\subsubsection{Experimental Setup}
In this research question, we aim to investigate whether our constructed high quality dataset can benefit the downstream APU task. 
As introduced previously, Tang et al.~\cite{tang2021using} proposed a dataset of comment-edit pairs based on their matching based approach. 
However, due to the limitations of their manually defined rules, their dataset contains significant invalid comment-edit pairs (more than 40\% according to our preliminary study). 
These invalid pairs introduce considerable noise that could negatively affect model training for the APU task. 
Therefore, we conducted a cross-dataset evaluation between different models. 
In particular, we prepared two datasets for this RQ: (1) \textbf{Tang-Data:} we apply Tang's matching method on our full-scale dataset to obtain comment-edit pairs, finally 17,411 comment-edit pairs are extracted, denoted as Tang-Data. 
Notably, Tang-Data is smaller than the dataset presented in their paper, this is because we only consider single code block edit in this study, while Tang et al.~\cite{tang2021using} collected edits with multi code blocks in their original paper.
(2) \textbf{HQ-Data:} This is the high-quality dataset constructed by \textsc{Soup}$_p$, the information of which is displayed in the last row of Table~\ref{tab:data_vcp}.
We split Tang-Data into train/validtion/test set by the ratio of 8:1:1, the details of these two datasets are displayed in Table~\ref{tab:data_rq3}. 
We perform a cross dataset evaluation for this research question, the models trained with HQ-Data's train set are also evaluated on Tang-Data's test set, and vice versa.

\begin{table}
\footnotesize
  \caption{Information on Datasets Constructed by Different Methods}
  \label{tab:data_rq3}
  \centering
  \begin{tabular}{lccc}
    \toprule
    \textbf{Dataset} & \textbf{Train} & \textbf{Validation} & \textbf{Test} \\
    \midrule
    Tang-Data & 13,929 & 1,741 & 1741 \\
    \midrule
    HQ-Data & 62,664 & 7,827 & 7,826 \\
  \bottomrule
\end{tabular}
\end{table}

\subsubsection{Evaluation Metrics and Baselines}
\zp{We reuse the \textit{EM} evaluation metric for this research question. 
We fine-tuned CoditT5 and CodeLlama with the train sets of above two datasets respectively, denoted as \textsc{Soup}-CoditT5 and \textsc{Soup}-CodeLlama. 
The models with their best performance on the validation set are chosen for test set evaluation.} 
\subsubsection{Experimental Results}
From Table~\ref{tab:rq3_results}, it can be seen that: (1) The models trained with our HQ-Data outperform the models trained with Tang-Data. 
For example, \textsc{Soup}-CoditT5 trained with HQ-Data achieves 25.9\% for \textit{EM} evaluation metric on its test set, while \textsc{Soup}-CoditT5 trained with Tang-Data only achieves 9.2\% \textit{EM} on its test set. 
This is because HQ-Data has its advantage over Tang-Data in terms of data quantity, such as our HQ-Data contains more than 62K data samples for training, while Tang-Data only contains 13K training samples, which misses a large number of valid pairs. 
(2) By directly applying models trained with our HQ-Data to Tang-Data's test set, we can still achieve a far better performance. 
For example, \textsc{Soup}-CoditT5 trained with HQ-Data achieve 15.1\% \textit{EM} on Tang-Data's test set, even better than the models trained with Tang-Data itself (e.g., \textit{EM} of 9.2\%). 
This is because our HQ-Data has its advantage over HQ-Data in terms of data quantity and data validity, according to evaluation results reported in RQ1, Tang-Data contains only 57.1\% of valid data, which means the dataset contains nearly half of the noise. 
(3) The models trained with Tang-Data can hardly adapt to our HQ-Data's test set. 
For example, when applying \textsc{Soup}-CoditT5 trained with Tang-Data to our HQ-Data's test set, the \textit{EM} significantly dropped from 9.2\% to 5.5\%. 
This is reasonable because our constructed HQ-Data covers wider usage scenarios, e.g., the valid comment-edit pairs that no code terms are mentioned or modified. 
Models trained with Tang-Data hardly learn these features and are unable to generate correct edits on these unseen samples. 

\renewcommand{\arraystretch}{1.3} 
\begin{table}
\caption{The Results of Cross Dataset Evaluation}
\footnotesize
\centering
\begin{tabular}{cccc}
\toprule
\multicolumn{2}{c}{\textbf{Approach}} & \textbf{Tang-Data} \textit{Test} & \textbf{HQ-Data} \textit{Test} \\ 
\midrule
Model & \textit{Train} & EM & EM \\ 
\midrule
\multirow{2}{*}{\textsc{Soup}-CoditT5} & Tang-Data & 9.2\% & 5.5\% \\ 
& HQ-Data & 15.1\% & \textbf{25.9}\% \\ 
\cline{1-4}
\multirow{2}{*}{\textsc{Soup}-CodeLlama} & Tang-Data & 24.8\% & 21.4\% \\ 
& HQ-Data & \textbf{25.6}\% & 25.6\% \\ 
\bottomrule
\end{tabular}
\label{tab:rq3_results}
\end{table}
\renewcommand{\arraystretch}{1}

\find{\textbf{Answer to RQ3: To what extent does the dataset influence model performance in the APU task?}
Our constructed dataset is superior to the dataset constructed by Tang's method in terms of quantity, validity and variety, which is beneficial to improving the performance of various models on the APU task in different scenarios.}

\subsection{RQ4: In-the-Wild Evaluation} 
\subsubsection{Experimental Setup}
\zp{The final goal of our \textsc{Soup} framework is to help Stack Overflow users and/or developers automatically update answer post code snippets based on their comments, in this research question, we perform an in-the-wild evaluation to evaluate the effectiveness of our \textsc{Soup} for updating code snippets in real world Stack Overflow posts. 
On Stack Overflow, users can edit a post when they want to improve the post quality, such as fixing typos or code related issues. 
After edits are suggested, it will be peer-reviewed by several senior Stack Overflow users whether the edits should be accepted or not. 
Once accepted, the posts will be updated with newly posted edits. 
}

\zp{To perform this in-the-wild evaluation, we collected 50 $\langle$code snippet, comment$\rangle$ pairs from Stack Overflow answer posts. 
We first set a time window of one year (from February 2023 to February 2024) to focus on recently raised comments, as they are more likely to remain unaddressed. 
After that, we filtered comments containing specific keywords (e.g., \textit{deprecate}, \textit{outdate}, \textit{error}, \textit{issue}), as these comments often point out weaknesses within code snippets and indicate a need for timely updates. 
We matched these comments with their code snippets, initially identifying 2175 candidate pairs. 
Due to the presence of noisy samples (e.g., comments without editing purposes), we further manually reviewed these pairs and only retained posts that needed updating. 
Considering the time and resource limitations, we finished the in-the-wild dataset construction when the sample size reached 50. 
For a given $\langle$code snippet, comment$\rangle$ pair, we then applied our \textsc{Soup}$_u$ model to generate the updated code snippet based on its comment. 
Two authors then posted these generated edits to the original posts for updating. 
}

\begin{figure}[]
\centering
\includegraphics[width = 0.99\linewidth]{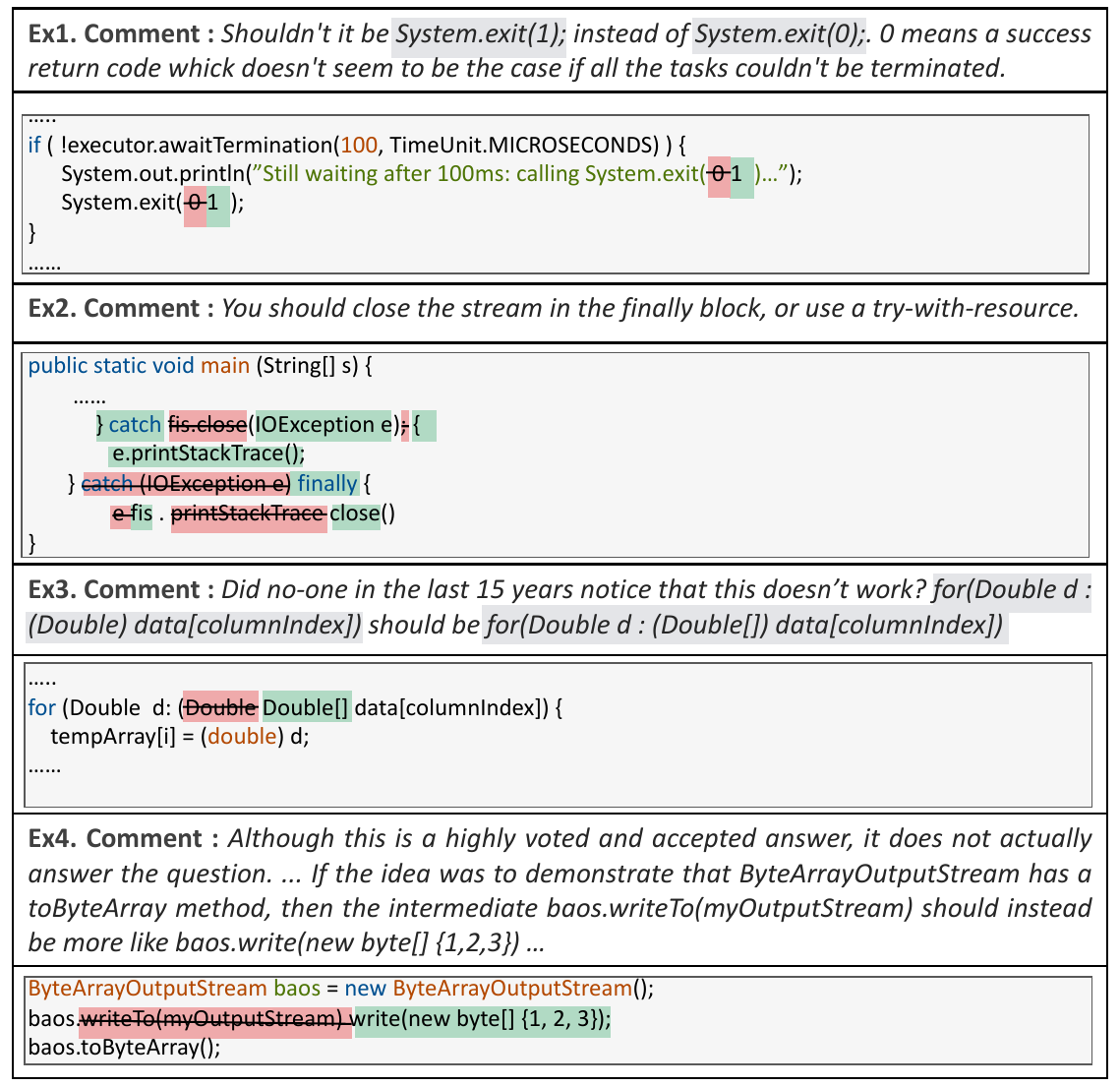}
\caption{In-the-Wild Evaluation Examples}
\vspace{-0.3cm}
\label{fig:in-the-wild}
\end{figure}

\subsubsection{Experimental Results}
\zp{We submitted 50 edit requests generated by our \textsc{Soup}$_u$ to their corresponding Stack Overflow posts and 21 of them have already been confirmed and accepted by Stack Overflow senior developers and maintainers. 
To avoid subjective bias, the developers were unaware of the edits were generated by our approach. 
Fig.~\ref{fig:in-the-wild} demonstrates three accepted examples (i.e., Ex1, Ex2, Ex3) and one rejected example (i.e., Ex4) from our in-the-wild evaluation. For the sake of double-blind reviewing, we hide the details of the \textit{PostId}. 
}

\textbf{Ex.1} 
demonstrates an accept answer post that obtains 158 votes, the latest edits occurred a decade ago (July 2012), however, a developer recently (February 2024) pointed out that it should use \texttt{exit(1)} instead of  \texttt{exit(0)}, because ``\textit{$0$ means a success return code}'', our approach successfully fix this problem with editing. 
Before our updating, this error has existed in Stack Overflow for more than 12 years, which can mislead developers in their daily development. 
\textbf{Ex.2} shows another edits verified by developers. 
The commenter mentioned ``\textit{You should close the stream in the finally block}'', our model successfully identifies \texttt{fis} in the \texttt{try} block and moves it to the \texttt{finally} block. 
Notably, no code terms are mentioned in this comment and our model can capture the semantic meaning of the comment and generate correct edits successfully. 
In \textbf{Ex.3}, multiple code terms are mentioned and our model successfully captures tiny differences for updating.

Our model does not always generate correct edits, one common failed situation is that the comments are too complex and/or do not provide valuable editing information (e.g., just posting error messages when reusing this code).
\textbf{Ex.4} demonstrates a rejected suggestion. 
In this case, our model wrongly updated this code snippet because the edits were mentioned as an example by the commenter, and SO maintainers rejected our edits because ``the edit does not improve the quality of the post''

\find{\textbf{Answer to RQ4: How acceptable are our updated SO posts in real-world scenarios?}
21 out of the 50 suggested edits generated by Soup were accepted, with an acceptance rate of 42\%, further demonstrating its practical value.}

\section{Discussion}
\label{sec:dis}
\subsection{Data Leakage Issues}
Our research examined three LLMs: CoditT5, CodeLlama, and ChatGPT. CoditT5, based on the CodeT5 model and pre-trained on the CodeSearchNet~\cite{husain2019codesearchnet} dataset from GitHub, was fine-tuned for comment updating, bug fixing, and automated code review without using SO data. 
CodeLlama, built on Llama2, incorporated StackExchange data in its pre-training. 
Since ChatGPT is a closed-source model, its use of SO data is unclear. 
However, the low performance of CodeLlama and ChatGPT in our study (EM scores of 0.2\% and 2.1\%, respectively) suggests they may not utilize the SO editing dataset in their model training or fine-tuning. 

\subsection{Ethical Issues}
As we utilize the SO dataset and GenAI in this study, there are potential ethical issues related to data privacy, intellectual property, and forum quality, as well as SO policy compliance. 
SO dataset is publicly available, we used the data solely for research without revealing user identities, we also aligned with SO's terms for non-commercial use. 
In terms of the quality concerns, which were addressed through careful review by experienced SO users and four authors and a limited experiment scale (50 cases). 
Despite SO's current policy against GenAI-generated content, our research aims to enhance SO post quality through LLMs, strictly for research. 
Overall, we argue the ethical issues of this study are limited. 

\subsection{LLMs versus Stack Overflow}
Although LLMs have made significant progress in code-related tasks, SO still provides valuable insights that current LLMs (e.g., ChatGPT) cannot fully replicate. 
On one hand, SO contains a wide range of practical problems that may exceed the capabilities of LLMs, for example, ChatGPT only achieved 2.1\% EM score in our APU task, even our model can only achieve 25.6\% EM score, which means LLMs still have a large room for improvement. On the other hand, the quality of the LLMs’ generated content still varies significantly. For example, the AI-generated answers can be false or misleading~\cite{zhong2024can}, while SO answers are more trustworthy since they have been voted on and peer-reviewed by community members and experts. Therefore, how to combine LLMs with SO to better help developers for solving their daily technical problems will be an interesting research direction in the future.

\section{Related Work}
\label{sec:related}
\subsection{Mining Stack Overflow Posts}
Currently, many studies have been conducted on Stack Overflow, including post recommendation~\cite{gao2020technical, cai2019biker, cai2019answerbot, wang2021automatic}, query reformulation~\cite{cao2021automated, gao2020generating, gao2021code2que}, and content quality analysis~\cite{zhang2019empirical, ragkhitwetsagul2019toxic, wang2018users, mai2024human, bi2021mining}. 
Zhang et al.~\cite{zhang2019empirical} found that 58.4\% of outdated SO answers were obsolete upon posting, with only 20.5\% updated upon discovery. 
Ragkhitwetsagul et al.\cite{ragkhitwetsagul2019toxic} reported that 66\% of users face issues with reused SO answers, such as mismatches and bugs. 
Wang et al.~\cite{wang2018users} linked SO badge incentives to minor edits, which are often reverted. 
Baltes et al.~\cite{baltes2018sotorrent, baltes2019sotorrent} created Sotorrent dataset to refine SO post edit history analysis. 
Diamantopoulos et al.~\cite{diamantopoulos2019towards} mined common edit patterns from Sotorrent. Adaji et al.~\cite{adaji2016modelling} found that high-quality answers usually contain more comments. 
Zhang et al.~\cite{zhang2019reading} categorized SO comments, noting their informativeness for answer optimization. 
Son et al.~\cite{soni2019analyzing} found that only 4.6\% of comments led to updates, with 27.5\% of update requests ignored. Tang et al.~\cite{tang2021using} developed a method to match comments with edits. While these studies highlight SO post issues and the value of comments for code optimization, they lack effective strategies for using comments to enhance SO posts. Their work provides a good research motivation for the APU task we proposed.

\subsection{Code Maintenance \& Code Review}
Datasets for code maintenance, often derived from programming competitions like Marmoset~\cite{spacco2005software}, IntroClass~\cite{le2015manybugs}, QuixBugs~\cite{lin2017quixbugs}, Codeflaws~\cite{tan2017codeflaws}, or open-source projects such as iBugs~\cite{dallmeier2007extraction}, Defects4j~\cite{just2014defects4j}, BugsJS~\cite{gyimesi2019bugsjs}, are pivotal for bug fixing tasks~\cite{li2024empirically, yang2024federated, qiu2021deep}. 
However, these datasets are typically limited in scale or quality due to manual construction or heuristic-based methods. 
Our work introduces a novel LLM-based framework, \textsc{Soup}$_p$, which automates dataset construction, yielding a dataset that surpasses existing ones in both quality and scale.


Code review datasets, on the other hand, focus on method-level changes and are used to generate revised code from comments and existing code, aiming to reduce maintenance costs. Unlike the APU task, which deals with fine-grained, often uncompileable code snippets~\cite{terragni2016csnippex,subramanian2013making,yang2016query}, code review datasets are not directly applicable to the APU task due to differing data characteristics. Nevertheless, successful model practices in code review, such as the Transforme~\cite{vaswani2017attention} model by Tufano et al.~\cite{tufano2021towards}, T5's~\cite{raffel2020exploring} performance evaluation~\cite{tufano2022using}, and the CoditT5 model by Zhang et al.~\cite{zhang2022coditt5}, provide valuable insights. Moreover, Zhao et al.~\cite{zhao2023right} have explored the efficacy of various large models, including ChatGPT, CodeLlama~\cite{roziere2023code}, and CodeReviewer~\cite{li2022automating}, in code review, identifying optimal prompts and models. While these LLMs have shown promise in code review, their applicability to the APU task remains untested, which our paper addresses. We are confident that our APU dataset will contribute to code maintenance and code review at a fine-grained level, and we will further investigate the applicability of the APU dataset in other domains in the future.

\section{Threats to Validity}
\label{sec:threats}

\noindent\textbf{Threats to internal validity.} The use of manually curated datasets in the VCP task introduces a risk of bias. We addressed this by engaging multiple annotators to reduce subjectivity in the annotation process. Additionally, the preliminary dataset analysis by the first author may carry inherent bias, which we mitigated by validating with objective datasets in RQ1 and RQ3, confirming the consistency of our results. The potential bias in manual error example selection was countered through stratified sampling, ensuring a more objective representation of results.

\noindent\textbf{Threats to external validity.} Our study's focus on Java snippets from Stack Overflow could limit broader applicability. To combat this limitation, we highlight the adaptability of our approach to other platforms and languages, with plans to explore this in future work.


\noindent\textbf{Threats to construct validity.} As our manual analysis section shows, those code snippets that do not precisely match the true code snippets can also be reasonable. The exact match metric merely reflects a lower bound of the model's ability to generate meaningful code snippets. Furthermore, those code snippets with high \textit{CodeBLEU} scores might also be incorrect. Since most code snippets from Stack Overflow are difficult to run directly~\cite{terragni2016csnippex,subramanian2013making,yang2016query}, implementing an automatic functional equivalence assessment is very challenging. Therefore, we further analyzed 50 randomly selected error examples to further verify the correctness of the code generated by \textsc{Soup}.


\section{Conclusion}
\label{sec:con}
In this work, we introduced \textsc{Soup}, a LLM-based framework to automatically update posts in Stack Overflow. 
Our key contributions are twofold: firstly, we constructed a high quality dataset of 78,317 valid comment-edit pairs. 
Secondly, we proposed a novel framework \textsc{Soup} to perform SO post updating, the extensive experiments show the superiority of our approach over a set of baselines. 
Our work first attempts to automatically update knowledge on platforms such as Stack Overflow, aiming to improve code quality and reliability for SO users and developers.

\section*{Acknowledgment}
This research is supported by the Starry Night Science Fund of Zhejiang University Shanghai Institute for Advanced Study, Grant No. SN-ZJU-SIAS-001. 
This research is supported by the National Key Research and Development Program of China (No. 2021YFB2701102). 
This research is partially supported by the Shanghai Sailing Program (23YF1446900) and the National Science Foundation of China (No. 62202341, No. 62302430, No.62372398, No.72342025, and U20A20173). 
This research is partially supported by the Ningbo Natural Science Foundation (No. 2023J292) and Zhejiang Provincial Natural Science Foundation of China (No. LQ24F020017). 
This research was also supported by the advanced computing resources provided by the Supercomputing Center of Hangzhou City University. 
The authors would like to thank the reviewers for their insightful and constructive feedback. 



\balance
\bibliographystyle{IEEEtran}
\bibliography{samples}

\begin{thebibliography}{10}
\providecommand{\url}[1]{#1}
\csname url@samestyle\endcsname
\providecommand{\newblock}{\relax}
\providecommand{\bibinfo}[2]{#2}
\providecommand{\BIBentrySTDinterwordspacing}{\spaceskip=0pt\relax}
\providecommand{\BIBentryALTinterwordstretchfactor}{4}
\providecommand{\BIBentryALTinterwordspacing}{\spaceskip=\fontdimen2\font plus
\BIBentryALTinterwordstretchfactor\fontdimen3\font minus \fontdimen4\font\relax}
\providecommand{\BIBforeignlanguage}[2]{{%
\expandafter\ifx\csname l@#1\endcsname\relax
\typeout{** WARNING: IEEEtran.bst: No hyphenation pattern has been}%
\typeout{** loaded for the language `#1'. Using the pattern for}%
\typeout{** the default language instead.}%
\else
\language=\csname l@#1\endcsname
\fi
#2}}
\providecommand{\BIBdecl}{\relax}
\BIBdecl

\bibitem{gao2020generating}
Z.~Gao, X.~Xia, J.~Grundy, D.~Lo, and Y.-F. Li, ``Generating question titles for stack overflow from mined code snippets,'' \emph{ACM Transactions on Software Engineering and Methodology (TOSEM)}, vol.~29, no.~4, pp. 1--37, 2020.

\bibitem{gao2023know}
Z.~Gao, X.~Xia, D.~Lo, J.~Grundy, X.~Zhang, and Z.~Xing, ``I know what you are searching for: Code snippet recommendation from stack overflow posts,'' \emph{ACM Transactions on Software Engineering and Methodology}, vol.~32, no.~3, pp. 1--42, 2023.

\bibitem{ragkhitwetsagul2019toxic}
C.~Ragkhitwetsagul, J.~Krinke, M.~Paixao, G.~Bianco, and R.~Oliveto, ``Toxic code snippets on stack overflow,'' \emph{IEEE Transactions on Software Engineering}, vol.~47, no.~3, pp. 560--581, 2019.

\bibitem{zhang2019empirical}
H.~Zhang, S.~Wang, T.-H. Chen, Y.~Zou, and A.~E. Hassan, ``An empirical study of obsolete answers on stack overflow,'' \emph{IEEE Transactions on Software Engineering}, vol.~47, no.~4, pp. 850--862, 2019.

\bibitem{nie2017data}
L.~Nie, X.~Wei, D.~Zhang, X.~Wang, Z.~Gao, and Y.~Yang, ``Data-driven answer selection in community qa systems,'' \emph{IEEE transactions on knowledge and data engineering}, vol.~29, no.~6, pp. 1186--1198, 2017.

\bibitem{gao2020technical}
Z.~Gao, X.~Xia, D.~Lo, and J.~Grundy, ``Technical q8a site answer recommendation via question boosting,'' \emph{ACM Transactions on Software Engineering and Methodology (TOSEM)}, vol.~30, no.~1, pp. 1--34, 2020.

\bibitem{intro_so}
PostURL, \url{https://stackoverflow.com/posts/comments/61093395/}.

\bibitem{tang2021using}
H.~Tang and S.~Nadi, ``On using stack overflow comment-edit pairs to recommend code maintenance changes,'' \emph{Empirical Software Engineering}, vol.~26, no.~4, p.~68, 2021.

\bibitem{touvron2023llama}
H.~Touvron, T.~Lavril, G.~Izacard, X.~Martinet, M.-A. Lachaux, T.~Lacroix, B.~Rozi{\`e}re, N.~Goyal, E.~Hambro, F.~Azhar \emph{et~al.}, ``Llama: Open and efficient foundation language models,'' \emph{arXiv preprint arXiv:2302.13971}, 2023.

\bibitem{wang2024software}
J.~Wang, Y.~Huang, C.~Chen, Z.~Liu, S.~Wang, and Q.~Wang, ``Software testing with large language models: Survey, landscape, and vision,'' \emph{IEEE Transactions on Software Engineering}, 2024.

\bibitem{wang2022self}
Y.~Wang, Y.~Kordi, S.~Mishra, A.~Liu, N.~A. Smith, D.~Khashabi, and H.~Hajishirzi, ``Self-instruct: Aligning language models with self-generated instructions,'' \emph{arXiv preprint arXiv:2212.10560}, 2022.

\bibitem{zhao2023survey}
W.~X. Zhao, K.~Zhou, J.~Li, T.~Tang, X.~Wang, Y.~Hou, Y.~Min, B.~Zhang, J.~Zhang, Z.~Dong \emph{et~al.}, ``A survey of large language models,'' \emph{arXiv preprint arXiv:2303.18223}, 2023.

\bibitem{mai2024human}
Y.~Mai, Z.~Gao, X.~Hu, L.~Bao, Y.~Liu, and J.~Sun, ``Are human rules necessary? generating reusable apis with cot reasoning and in-context learning,'' \emph{Proceedings of the ACM on Software Engineering}, vol.~1, no. FSE, pp. 2355--2377, 2024.

\bibitem{wang2024just}
H.~Wang, Z.~Gao, X.~Hu, D.~Lo, J.~Grundy, and X.~Wang, ``Just-in-time todo-missed commits detection,'' \emph{IEEE Transactions on Software Engineering}, 2024.

\bibitem{wang2024makes}
H.~Wang, Z.~Gao, T.~Bi, J.~Grundy, X.~Wang, M.~Wu, and X.~Yang, ``What makes a good todo comment?'' \emph{ACM Transactions on Software Engineering and Methodology}.

\bibitem{gao2021automating}
Z.~Gao, X.~Xia, D.~Lo, J.~Grundy, and T.~Zimmermann, ``Automating the removal of obsolete todo comments,'' in \emph{Proceedings of the 29th ACM Joint Meeting on European Software Engineering Conference and Symposium on the Foundations of Software Engineering}, 2021, pp. 218--229.

\bibitem{gao2024automating}
Z.~Gao, Y.~Su, X.~Hu, and X.~Xia, ``Automating todo-missed methods detection and patching,'' \emph{arXiv preprint arXiv:2405.06225}, 2024.

\bibitem{fried2022incoder}
D.~Fried, A.~Aghajanyan, J.~Lin, S.~Wang, E.~Wallace, F.~Shi, R.~Zhong, W.-t. Yih, L.~Zettlemoyer, and M.~Lewis, ``Incoder: A generative model for code infilling and synthesis,'' \emph{arXiv preprint arXiv:2204.05999}, 2022.

\bibitem{jiang2023selfevolve}
S.~Jiang, Y.~Wang, and Y.~Wang, ``Selfevolve: A code evolution framework via large language models,'' \emph{arXiv preprint arXiv:2306.02907}, 2023.

\bibitem{lu2021codexglue}
S.~Lu, D.~Guo, S.~Ren, J.~Huang, A.~Svyatkovskiy, A.~Blanco, C.~Clement, D.~Drain, D.~Jiang, D.~Tang \emph{et~al.}, ``Codexglue: A machine learning benchmark dataset for code understanding and generation,'' \emph{arXiv preprint arXiv:2102.04664}, 2021.

\bibitem{zheng2023codegeex}
Q.~Zheng, X.~Xia, X.~Zou, Y.~Dong, S.~Wang, Y.~Xue, Z.~Wang, L.~Shen, A.~Wang, Y.~Li \emph{et~al.}, ``Codegeex: A pre-trained model for code generation with multilingual evaluations on humaneval-x,'' \emph{arXiv preprint arXiv:2303.17568}, 2023.

\bibitem{yan2023closer}
D.~Yan, Z.~Gao, and Z.~Liu, ``A closer look at different difficulty levels code generation abilities of chatgpt,'' in \emph{2023 38th IEEE/ACM International Conference on Automated Software Engineering (ASE)}.\hskip 1em plus 0.5em minus 0.4em\relax IEEE, 2023, pp. 1887--1898.

\bibitem{dai2024mpcoder}
Z.~Dai, C.~Yao, W.~Han, Y.~Yuan, Z.~Gao, and J.~Chen, ``Mpcoder: Multi-user personalized code generator with explicit and implicit style representation learning,'' \emph{arXiv preprint arXiv:2406.17255}, 2024.

\bibitem{xue2024selfpico}
Z.~Xue, Z.~Gao, S.~Wang, X.~Hu, X.~Xia, and S.~Li, ``Selfpico: Self-guided partial code execution with llms,'' \emph{arXiv preprint arXiv:2407.16974}, 2024.

\bibitem{replicate}
\BIBentryALTinterwordspacing
Soup, ``Our replicate package,'' 2024. [Online]. Available: \url{https://doi.org/10.6084/m9.figshare.25465282}
\BIBentrySTDinterwordspacing

\bibitem{soni2019analyzing}
A.~Soni and S.~Nadi, ``Analyzing comment-induced updates on stack overflow,'' in \emph{2019 IEEE/ACM 16th International Conference on Mining Software Repositories (MSR)}.\hskip 1em plus 0.5em minus 0.4em\relax IEEE, 2019, pp. 220--224.

\bibitem{rosner2006fundamentals}
B.~A. Rosner \emph{et~al.}, \emph{Fundamentals of biostatistics}.\hskip 1em plus 0.5em minus 0.4em\relax Thomson-Brooks/Cole Belmont, CA, 2006, vol.~6.

\bibitem{baltes2018sotorrent}
S.~Baltes, L.~Dumani, C.~Treude, and S.~Diehl, ``Sotorrent: reconstructing and analyzing the evolution of stack overflow posts,'' in \emph{Proceedings of the 15th international conference on mining software repositories}, 2018, pp. 319--330.

\bibitem{cohen1968weighted}
J.~Cohen, ``Weighted kappa: Nominal scale agreement provision for scaled disagreement or partial credit.'' \emph{Psychological bulletin}, vol.~70, no.~4, p. 213, 1968.

\bibitem{landis1977measurement}
J.~R. Landis and G.~G. Koch, ``The measurement of observer agreement for categorical data,'' \emph{biometrics}, pp. 159--174, 1977.

\bibitem{krawczyk2016learning}
B.~Krawczyk, ``Learning from imbalanced data: open challenges and future directions,'' \emph{Progress in artificial intelligence}, vol.~5, no.~4, pp. 221--232, 2016.

\bibitem{lester2021power}
B.~Lester, R.~Al-Rfou, and N.~Constant, ``The power of scale for parameter-efficient prompt tuning,'' \emph{arXiv preprint arXiv:2104.08691}, 2021.

\bibitem{yang2024automatic}
S.~Yang, X.~Chen, K.~Liu, G.~Yang, and C.~Yu, ``Automatic bi-modal question title generation for stack overflow with prompt learning,'' \emph{arXiv preprint arXiv:2403.03677}, 2024.

\bibitem{huang2022prompt}
Q.~Huang, Z.~Yuan, Z.~Xing, X.~Xu, L.~Zhu, and Q.~Lu, ``Prompt-tuned code language model as a neural knowledge base for type inference in statically-typed partial code,'' in \emph{Proceedings of the 37th IEEE/ACM International Conference on Automated Software Engineering}, 2022, pp. 1--13.

\bibitem{feng2024prompting}
S.~Feng and C.~Chen, ``Prompting is all you need: Automated android bug replay with large language models,'' in \emph{Proceedings of the 46th IEEE/ACM International Conference on Software Engineering}, 2024, pp. 1--13.

\bibitem{wei2022chain}
J.~Wei, X.~Wang, D.~Schuurmans, M.~Bosma, F.~Xia, E.~Chi, Q.~V. Le, D.~Zhou \emph{et~al.}, ``Chain-of-thought prompting elicits reasoning in large language models,'' \emph{Advances in neural information processing systems}, vol.~35, pp. 24\,824--24\,837, 2022.

\bibitem{roziere2023code}
B.~Roziere, J.~Gehring, F.~Gloeckle, S.~Sootla, I.~Gat, X.~E. Tan, Y.~Adi, J.~Liu, T.~Remez, J.~Rapin \emph{et~al.}, ``Code llama: Open foundation models for code,'' \emph{arXiv preprint arXiv:2308.12950}, 2023.

\bibitem{hu2021lora}
E.~J. Hu, Y.~Shen, P.~Wallis, Z.~Allen-Zhu, Y.~Li, S.~Wang, L.~Wang, and W.~Chen, ``Lora: Low-rank adaptation of large language models,'' \emph{arXiv preprint arXiv:2106.09685}, 2021.

\bibitem{ren2020codebleu}
S.~Ren, D.~Guo, S.~Lu, L.~Zhou, S.~Liu, D.~Tang, N.~Sundaresan, M.~Zhou, A.~Blanco, and S.~Ma, ``Codebleu: a method for automatic evaluation of code synthesis,'' \emph{arXiv preprint arXiv:2009.10297}, 2020.

\bibitem{zhang2022coditt5}
J.~Zhang, S.~Panthaplackel, P.~Nie, J.~J. Li, and M.~Gligoric, ``Coditt5: Pretraining for source code and natural language editing,'' in \emph{Proceedings of the 37th IEEE/ACM International Conference on Automated Software Engineering}, 2022, pp. 1--12.

\bibitem{wang2021codet5}
Y.~Wang, W.~Wang, S.~Joty, and S.~C. Hoi, ``Codet5: Identifier-aware unified pre-trained encoder-decoder models for code understanding and generation,'' \emph{arXiv preprint arXiv:2109.00859}, 2021.

\bibitem{panthaplackel2020learning}
S.~Panthaplackel, P.~Nie, M.~Gligoric, J.~J. Li, and R.~J. Mooney, ``Learning to update natural language comments based on code changes,'' \emph{arXiv preprint arXiv:2004.12169}, 2020.

\bibitem{tufano2019learning}
M.~Tufano, J.~Pantiuchina, C.~Watson, G.~Bavota, and D.~Poshyvanyk, ``On learning meaningful code changes via neural machine translation,'' in \emph{2019 IEEE/ACM 41st International Conference on Software Engineering (ICSE)}.\hskip 1em plus 0.5em minus 0.4em\relax IEEE, 2019, pp. 25--36.

\bibitem{tufano2022using}
R.~Tufano, S.~Masiero, A.~Mastropaolo, L.~Pascarella, D.~Poshyvanyk, and G.~Bavota, ``Using pre-trained models to boost code review automation,'' in \emph{Proceedings of the 44th international conference on software engineering}, 2022, pp. 2291--2302.

\bibitem{clement2020pymt5}
C.~B. Clement, D.~Drain, J.~Timcheck, A.~Svyatkovskiy, and N.~Sundaresan, ``Pymt5: multi-mode translation of natural language and python code with transformers,'' \emph{arXiv preprint arXiv:2010.03150}, 2020.

\bibitem{feng2020codebert}
Z.~Feng, D.~Guo, D.~Tang, N.~Duan, X.~Feng, M.~Gong, L.~Shou, B.~Qin, T.~Liu, D.~Jiang \emph{et~al.}, ``Codebert: A pre-trained model for programming and natural languages,'' \emph{arXiv preprint arXiv:2002.08155}, 2020.

\bibitem{chen2022codet}
B.~Chen, F.~Zhang, A.~Nguyen, D.~Zan, Z.~Lin, J.-G. Lou, and W.~Chen, ``Codet: Code generation with generated tests,'' \emph{arXiv preprint arXiv:2207.10397}, 2022.

\bibitem{sobania2023analysis}
D.~Sobania, M.~Briesch, C.~Hanna, and J.~Petke, ``An analysis of the automatic bug fixing performance of chatgpt,'' in \emph{2023 IEEE/ACM International Workshop on Automated Program Repair (APR)}.\hskip 1em plus 0.5em minus 0.4em\relax IEEE, 2023, pp. 23--30.

\bibitem{zhang2019reading}
H.~Zhang, S.~Wang, T.-H. Chen, and A.~E. Hassan, ``Reading answers on stack overflow: Not enough!'' \emph{IEEE Transactions on Software Engineering}, vol.~47, no.~11, pp. 2520--2533, 2019.

\bibitem{rq2_so}
PostURL, \url{https://stackoverflow.com/questions/46980301}.

\bibitem{husain2019codesearchnet}
H.~Husain, H.-H. Wu, T.~Gazit, M.~Allamanis, and M.~Brockschmidt, ``Codesearchnet challenge: Evaluating the state of semantic code search,'' \emph{arXiv preprint arXiv:1909.09436}, 2019.

\bibitem{zhong2024can}
L.~Zhong and Z.~Wang, ``Can llm replace stack overflow? a study on robustness and reliability of large language model code generation,'' in \emph{Proceedings of the AAAI Conference on Artificial Intelligence}, vol.~38, no.~19, 2024, pp. 21\,841--21\,849.

\bibitem{cai2019biker}
L.~Cai, H.~Wang, Q.~Huang, X.~Xia, Z.~Xing, and D.~Lo, ``Biker: a tool for bi-information source based api method recommendation,'' in \emph{Proceedings of the 2019 27th ACM Joint Meeting on European Software Engineering Conference and Symposium on the Foundations of Software Engineering}, 2019, pp. 1075--1079.

\bibitem{cai2019answerbot}
L.~Cai, H.~Wang, B.~Xu, Q.~Huang, X.~Xia, D.~Lo, and Z.~Xing, ``Answerbot: an answer summary generation tool based on stack overflow,'' in \emph{Proceedings of the 2019 27th ACM Joint Meeting on European Software Engineering Conference and Symposium on the Foundations of Software Engineering}, 2019, pp. 1134--1138.

\bibitem{wang2021automatic}
H.~Wang, X.~Xia, D.~Lo, J.~Grundy, and X.~Wang, ``Automatic solution summarization for crash bugs,'' in \emph{2021 IEEE/ACM 43rd International Conference on Software Engineering (ICSE)}.\hskip 1em plus 0.5em minus 0.4em\relax IEEE, 2021, pp. 1286--1297.

\bibitem{cao2021automated}
K.~Cao, C.~Chen, S.~Baltes, C.~Treude, and X.~Chen, ``Automated query reformulation for efficient search based on query logs from stack overflow,'' in \emph{2021 IEEE/ACM 43rd International Conference on Software Engineering (ICSE)}.\hskip 1em plus 0.5em minus 0.4em\relax IEEE, 2021, pp. 1273--1285.

\bibitem{gao2021code2que}
Z.~Gao, X.~Xia, D.~Lo, J.~Grundy, and Y.-F. Li, ``Code2que: A tool for improving question titles from mined code snippets in stack overflow,'' in \emph{Proceedings of the 29th ACM Joint Meeting on European Software Engineering Conference and Symposium on the Foundations of Software Engineering}, 2021, pp. 1525--1529.

\bibitem{wang2018users}
S.~Wang, T.-H. Chen, and A.~E. Hassan, ``How do users revise answers on technical q\&a websites? a case study on stack overflow,'' \emph{IEEE Transactions on Software Engineering}, vol.~46, no.~9, pp. 1024--1038, 2018.

\bibitem{bi2021mining}
T.~Bi, P.~Liang, A.~Tang, and X.~Xia, ``Mining architecture tactics and quality attributes knowledge in stack overflow,'' \emph{Journal of Systems and Software}, vol. 180, p. 111005, 2021.

\bibitem{baltes2019sotorrent}
S.~Baltes, C.~Treude, and S.~Diehl, ``Sotorrent: Studying the origin, evolution, and usage of stack overflow code snippets,'' in \emph{2019 IEEE/ACM 16th International Conference on Mining Software Repositories (MSR)}.\hskip 1em plus 0.5em minus 0.4em\relax IEEE, 2019, pp. 191--194.

\bibitem{diamantopoulos2019towards}
T.~Diamantopoulos, M.~I. Sifaki, and A.~Symeonidis, ``Towards mining answer edits to extract evolution patterns in stack overflow,'' in \emph{2019 IEEE/ACM 16th International Conference on Mining Software Repositories (MSR)}.\hskip 1em plus 0.5em minus 0.4em\relax IEEE, 2019, pp. 215--219.

\bibitem{adaji2016modelling}
I.~Adaji and J.~Vassileva, ``Modelling user collaboration in social networks using edits and comments,'' in \emph{Proceedings of the 2016 Conference on User Modeling Adaptation and Personalization}, 2016, pp. 111--114.

\bibitem{spacco2005software}
J.~Spacco, J.~Strecker, D.~Hovemeyer, and W.~Pugh, ``Software repository mining with marmoset: An automated programming project snapshot and testing system,'' in \emph{Proceedings of the 2005 international workshop on Mining software repositories}, 2005, pp. 1--5.

\bibitem{le2015manybugs}
C.~Le~Goues, N.~Holtschulte, E.~K. Smith, Y.~Brun, P.~Devanbu, S.~Forrest, and W.~Weimer, ``The manybugs and introclass benchmarks for automated repair of c programs,'' \emph{IEEE Transactions on Software Engineering}, vol.~41, no.~12, pp. 1236--1256, 2015.

\bibitem{lin2017quixbugs}
D.~Lin, J.~Koppel, A.~Chen, and A.~Solar-Lezama, ``Quixbugs: A multi-lingual program repair benchmark set based on the quixey challenge,'' in \emph{Proceedings Companion of the 2017 ACM SIGPLAN international conference on systems, programming, languages, and applications: software for humanity}, 2017, pp. 55--56.

\bibitem{tan2017codeflaws}
S.~H. Tan, J.~Yi, S.~Mechtaev, A.~Roychoudhury \emph{et~al.}, ``Codeflaws: a programming competition benchmark for evaluating automated program repair tools,'' in \emph{2017 IEEE/ACM 39th International Conference on Software Engineering Companion (ICSE-C)}.\hskip 1em plus 0.5em minus 0.4em\relax IEEE, 2017, pp. 180--182.

\bibitem{dallmeier2007extraction}
V.~Dallmeier and T.~Zimmermann, ``Extraction of bug localization benchmarks from history,'' in \emph{Proceedings of the 22nd IEEE/ACM international conference on automated software engineering}, 2007, pp. 433--436.

\bibitem{just2014defects4j}
R.~Just, D.~Jalali, and M.~D. Ernst, ``Defects4j: A database of existing faults to enable controlled testing studies for java programs,'' in \emph{Proceedings of the 2014 international symposium on software testing and analysis}, 2014, pp. 437--440.

\bibitem{gyimesi2019bugsjs}
P.~Gyimesi, B.~Vancsics, A.~Stocco, D.~Mazinanian, A.~Besz{\'e}des, R.~Ferenc, and A.~Mesbah, ``Bugsjs: a benchmark of javascript bugs,'' in \emph{2019 12th IEEE Conference on Software Testing, Validation and Verification (ICST)}.\hskip 1em plus 0.5em minus 0.4em\relax IEEE, 2019, pp. 90--101.

\bibitem{li2024empirically}
Z.~Li, M.~Pan, Y.~Pei, T.~Zhang, L.~Wang, and X.~Li, ``Empirically revisiting and enhancing automatic classification of bug and non-bug issues,'' \emph{Frontiers of Computer Science}, vol.~18, no.~5, p. 185207, 2024.

\bibitem{yang2024federated}
Y.~Yang, X.~Hu, Z.~Gao, J.~Chen, C.~Ni, X.~Xia, and D.~Lo, ``Federated learning for software engineering: A case study of code clone detection and defect prediction,'' \emph{IEEE Transactions on Software Engineering}, 2024.

\bibitem{qiu2021deep}
F.~Qiu, Z.~Gao, X.~Xia, D.~Lo, J.~Grundy, and X.~Wang, ``Deep just-in-time defect localization,'' \emph{IEEE Transactions on Software Engineering}, vol.~48, no.~12, pp. 5068--5086, 2021.

\bibitem{terragni2016csnippex}
V.~Terragni, Y.~Liu, and S.-C. Cheung, ``Csnippex: automated synthesis of compilable code snippets from q\&a sites,'' in \emph{Proceedings of the 25th international symposium on software testing and analysis}, 2016, pp. 118--129.

\bibitem{subramanian2013making}
S.~Subramanian and R.~Holmes, ``Making sense of online code snippets,'' in \emph{2013 10th Working Conference on Mining Software Repositories (MSR)}.\hskip 1em plus 0.5em minus 0.4em\relax IEEE, 2013, pp. 85--88.

\bibitem{yang2016query}
D.~Yang, A.~Hussain, and C.~V. Lopes, ``From query to usable code: an analysis of stack overflow code snippets,'' in \emph{Proceedings of the 13th International Conference on Mining Software Repositories}, 2016, pp. 391--402.

\bibitem{vaswani2017attention}
A.~Vaswani, N.~Shazeer, N.~Parmar, J.~Uszkoreit, L.~Jones, A.~N. Gomez, {\L}.~Kaiser, and I.~Polosukhin, ``Attention is all you need,'' \emph{Advances in neural information processing systems}, vol.~30, 2017.

\bibitem{tufano2021towards}
R.~Tufano, L.~Pascarella, M.~Tufano, D.~Poshyvanyk, and G.~Bavota, ``Towards automating code review activities,'' in \emph{2021 IEEE/ACM 43rd International Conference on Software Engineering (ICSE)}.\hskip 1em plus 0.5em minus 0.4em\relax IEEE, 2021, pp. 163--174.

\bibitem{raffel2020exploring}
C.~Raffel, N.~Shazeer, A.~Roberts, K.~Lee, S.~Narang, M.~Matena, Y.~Zhou, W.~Li, and P.~J. Liu, ``Exploring the limits of transfer learning with a unified text-to-text transformer,'' \emph{Journal of machine learning research}, vol.~21, no. 140, pp. 1--67, 2020.

\bibitem{zhao2023right}
Z.~Zhao, Z.~Xu, J.~Zhu, P.~Di, Y.~Yao, and X.~Ma, ``The right prompts for the job: Repair code-review defects with large language model,'' \emph{arXiv preprint arXiv:2312.17485}, 2023.

\bibitem{li2022automating}
Z.~Li, S.~Lu, D.~Guo, N.~Duan, S.~Jannu, G.~Jenks, D.~Majumder, J.~Green, A.~Svyatkovskiy, S.~Fu \emph{et~al.}, ``Automating code review activities by large-scale pre-training,'' in \emph{Proceedings of the 30th ACM Joint European Software Engineering Conference and Symposium on the Foundations of Software Engineering}, 2022, pp. 1035--1047.

\end{thebibliography}

\end{document}